\documentclass[prd,reprint,showpacs,superscriptaddress,bibnotes,floatfix]{revtex4-1}
\usepackage{natbib}
\usepackage{graphics}
\usepackage{graphicx}
\usepackage{multirow}
\usepackage{amsbsy}
\usepackage{mathrsfs}
\usepackage{footmisc}
\usepackage{hyperref}
\usepackage{amsmath}
\hypersetup{
 colorlinks = true,
  urlcolor = black,
  linkcolor=black,
  citecolor=black
}

\def\rrange{0.014^{+0.010}_{-0.011}}
\def\rul{0.036}
\def\rztop{0.46}
\def\rztopps{11}

\def\Adrange{4.4^{+0.8}_{-0.7}}
\def\Adcentval{4.4}

\def\Asul{1.4}

\def\Bdrange{1.49^{+0.13}_{-0.12}}

\def\rmlm{0.011}
\def\Asmlm{0.6}
\def\Admlm{4.4}
\def\Bsmlm{-3.0}
\def\asmlm{0.00}
\def\Bdmlm{1.5}
\def\admlm{-0.66}
\def\emlm{-0.11}

\def\chitwo{541.6}
\def\chitwoptesim{0.94}
\def\chioneptesim{0.22}

\def\Ad{A_\mathrm{d}}
\def\Adf{A_\mathrm{d,353}}
\def\As{A_\mathrm{sync}}
\def\Asf{A_\mathrm{sync,23}}

\def\Bd{\beta_\mathrm{d}}
\def\Bs{\beta_\mathrm{s}}

\def\ad{\alpha_\mathrm{d}}
\def\as{\alpha_\mathrm{s}}

\def\dd{\Delta_\mathrm{d}}

\def\bicep{BICEP}

\def\bicepone{{\sc BICEP1}}
\def\biceptwo{{\sc BICEP2}}
\def\bicepthree{{\sc BICEP3}}
\def\biceparray{{\sc BICEP} Array}

\def\keck{{\it Keck}}
\def\keckarray{{\it Keck Array}}

\def\planck{{\it Planck}} 

\def\wmap{WMAP}
\def\spt{SPT}

\def\sptpol{{\sc SPTpol}}

\def\spass{S-PASS}

\def\uksq{$\mu{\mathrm K^2}$}
\def\ukcmb{$\mu{\mathrm K}_{\mathrm{\mbox{\tiny\sc cmb}}}$}

\def\deg{^\circ}
\def\emode{$E$-mode}
\def\bmode{$B$-mode}

\def\clstar{\ell \left( \ell + 1 \right) C_\ell / 2 \pi}
\def\lcdm{$\Lambda$CDM}

\def\tp{$T \rightarrow P$}

\def\aap{Astron.\ Astrophys.}

\def\apj{Astrophys.\ J.}

\def\apjs{Astrophys.\ J.\ Suppl.\ Ser.}

\def\jcap{J.\ Cosmol.\ Astropart.\ Phys.}

\def\mnras{Mon.\ Not.\ R.\ Astron.\ Soc.}

\def\prd{Phys.\ Rev.\ D}

\def\prl{Phys.\ Rev.\ Lett.}

\begin{document}

\title{\bicep\ / \keck\ XIII:
Improved Constraints on Primordial Gravitational Waves using \planck, \wmap, and
\bicep/\keck\ Observations through the 2018 Observing Season}

\author{\bicep/\keck\ Collaboration: P.~A.~R.~Ade}
\affiliation{School of Physics and Astronomy, Cardiff University, Cardiff, CF24 3AA, United Kingdom}
\author{Z.~Ahmed}
\affiliation{Kavli Institute for Particle Astrophysics and Cosmology, SLAC National Accelerator Laboratory, 2575 Sand Hill Rd, Menlo Park, California 94025, USA}
\author{M.~Amiri}
\affiliation{Department of Physics and Astronomy, University of British Columbia, Vancouver, British Columbia, V6T 1Z1, Canada}
\author{D.~Barkats}
\affiliation{Center for Astrophysics, Harvard \& Smithsonian, Cambridge, MA 02138, U.S.A}
\author{R.~Basu Thakur}
\affiliation{Department of Physics, California Institute of Technology, Pasadena, California 91125, USA}
\author{C.~A.~Bischoff}
\affiliation{Department of Physics, University of Cincinnati, Cincinnati, Ohio 45221, USA}
\author{D.~Beck}
\affiliation{Kavli Institute for Particle Astrophysics and Cosmology, SLAC National Accelerator Laboratory, 2575 Sand Hill Rd, Menlo Park, California 94025, USA}
\affiliation{Department of Physics, Stanford University, Stanford, California 94305, USA}
\author{J.~J.~Bock}
\affiliation{Department of Physics, California Institute of Technology, Pasadena, California 91125, USA}
\affiliation{Jet Propulsion Laboratory, Pasadena, California 91109, USA}
\author{H.~Boenish}
\affiliation{Center for Astrophysics, Harvard \& Smithsonian, Cambridge, MA 02138, U.S.A}
\author{E.~Bullock}
\affiliation{Minnesota Institute for Astrophysics, University of Minnesota, Minneapolis, Minnesota 55455, USA}
\author{V.~Buza}
\affiliation{Kavli Institute for Cosmological Physics, University of Chicago, Chicago, IL 60637, USA}
\author{J.~R.~Cheshire IV}
\affiliation{Minnesota Institute for Astrophysics, University of Minnesota, Minneapolis, Minnesota 55455, USA}
\author{J.~Connors}
\affiliation{Center for Astrophysics, Harvard \& Smithsonian, Cambridge, MA 02138, U.S.A}
\author{J.~Cornelison}
\affiliation{Center for Astrophysics, Harvard \& Smithsonian, Cambridge, MA 02138, U.S.A}
\author{M.~Crumrine}
\affiliation{School of Physics and Astronomy, University of Minnesota, Minneapolis, Minnesota 55455, USA}
\author{A.~Cukierman}
\affiliation{Department of Physics, Stanford University, Stanford, California 94305, USA}
\affiliation{Kavli Institute for Particle Astrophysics and Cosmology, SLAC National Accelerator Laboratory, 2575 Sand Hill Rd, Menlo Park, California 94025, USA}
\author{E.~V.~Denison}
\affiliation{National Institute of Standards and Technology, Boulder, Colorado 80305, USA}
\author{M.~Dierickx}
\affiliation{Center for Astrophysics, Harvard \& Smithsonian, Cambridge, MA 02138, U.S.A}
\author{L.~Duband}
\affiliation{Service des Basses Temp\'{e}ratures, Commissariat \`{a} l'Energie Atomique, 38054 Grenoble, France}
\author{M.~Eiben}
\affiliation{Center for Astrophysics, Harvard \& Smithsonian, Cambridge, MA 02138, U.S.A}
\author{S.~Fatigoni}
\affiliation{Department of Physics and Astronomy, University of British Columbia, Vancouver, British Columbia, V6T 1Z1, Canada}
\author{J.~P.~Filippini}
\affiliation{Department of Physics, University of Illinois at Urbana-Champaign, Urbana, Illinois 61801, USA}
\affiliation{Department of Astronomy, University of Illinois at Urbana-Champaign, Urbana, Illinois 61801, USA}
\author{S.~Fliescher}
\affiliation{School of Physics and Astronomy, University of Minnesota, Minneapolis, Minnesota 55455, USA}
\author{N.~Goeckner-Wald}
\affiliation{Department of Physics, Stanford University, Stanford, California 94305, USA}
\author{D.~C.~Goldfinger}
\affiliation{Center for Astrophysics, Harvard \& Smithsonian, Cambridge, MA 02138, U.S.A}
\author{J.~Grayson}
\affiliation{Department of Physics, Stanford University, Stanford, California 94305, USA}
\author{P.~Grimes}
\affiliation{Center for Astrophysics, Harvard \& Smithsonian, Cambridge, MA 02138, U.S.A}
\author{G.~Hall}
\affiliation{School of Physics and Astronomy, University of Minnesota, Minneapolis, Minnesota 55455, USA}
\author{G. Halal}
\affiliation{Department of Physics, Stanford University, Stanford, California 94305, USA}
\author{M.~Halpern}
\affiliation{Department of Physics and Astronomy, University of British Columbia, Vancouver, British Columbia, V6T 1Z1, Canada}
\author{E.~Hand}
\affiliation{Department of Physics, University of Cincinnati, Cincinnati, Ohio 45221, USA}
\author{S.~Harrison}
\affiliation{Center for Astrophysics, Harvard \& Smithsonian, Cambridge, MA 02138, U.S.A}
\author{S. Henderson}
\affiliation{Kavli Institute for Particle Astrophysics and Cosmology, SLAC National Accelerator Laboratory, 2575 Sand Hill Rd, Menlo Park, California 94025, USA}
\author{S.~R.~Hildebrandt}
\affiliation{Department of Physics, California Institute of Technology, Pasadena, California 91125, USA}
\affiliation{Jet Propulsion Laboratory, Pasadena, California 91109, USA}
\author{G.~C.~Hilton}
\affiliation{National Institute of Standards and Technology, Boulder, Colorado 80305, USA}
\author{J.~Hubmayr}
\affiliation{National Institute of Standards and Technology, Boulder, Colorado 80305, USA}
\author{H.~Hui}
\affiliation{Department of Physics, California Institute of Technology, Pasadena, California 91125, USA}
\author{K.~D.~Irwin}
\affiliation{Department of Physics, Stanford University, Stanford, California 94305, USA}
\affiliation{Kavli Institute for Particle Astrophysics and Cosmology, SLAC National Accelerator Laboratory, 2575 Sand Hill Rd, Menlo Park, California 94025, USA}
\affiliation{National Institute of Standards and Technology, Boulder, Colorado 80305, USA}
\author{J.~Kang}
\affiliation{Department of Physics, Stanford University, Stanford, California 94305, USA}
\affiliation{Department of Physics, California Institute of Technology, Pasadena, California 91125, USA}
\author{K.~S.~Karkare}
\affiliation{Center for Astrophysics, Harvard \& Smithsonian, Cambridge, MA 02138, U.S.A}
\affiliation{Kavli Institute for Cosmological Physics, University of Chicago, Chicago, IL 60637, USA}
\author{E.~Karpel}
\affiliation{Department of Physics, Stanford University, Stanford, California 94305, USA}
\author{S.~Kefeli}
\affiliation{Department of Physics, California Institute of Technology, Pasadena, California 91125, USA}
\author{S.~A.~Kernasovskiy}
\affiliation{Department of Physics, Stanford University, Stanford, California 94305, USA}
\author{J.~M.~Kovac}
\affiliation{Center for Astrophysics, Harvard \& Smithsonian, Cambridge, MA 02138, U.S.A}
\affiliation{Department of Physics, Harvard University, Cambridge, MA 02138, USA}
\author{C.~L.~Kuo}
\affiliation{Department of Physics, Stanford University, Stanford, California 94305, USA}
\affiliation{Kavli Institute for Particle Astrophysics and Cosmology, SLAC National Accelerator Laboratory, 2575 Sand Hill Rd, Menlo Park, California 94025, USA}
\author{K.~Lau}
\affiliation{School of Physics and Astronomy, University of Minnesota, Minneapolis, Minnesota 55455, USA}
\author{E.~M.~Leitch}
\affiliation{Kavli Institute for Cosmological Physics, University of Chicago, Chicago, IL 60637, USA}
\author{A.~Lennox}
\affiliation{Department of Physics, University of Illinois at Urbana-Champaign, Urbana, Illinois 61801, USA}
\author{K.~G.~Megerian}
\affiliation{Jet Propulsion Laboratory, Pasadena, California 91109, USA}
\author{L.~Minutolo}
\affiliation{Department of Physics, California Institute of Technology, Pasadena, California 91125, USA}
\author{L.~Moncelsi}
\affiliation{Department of Physics, California Institute of Technology, Pasadena, California 91125, USA}
\author{Y. Nakato}
\affiliation{Department of Physics, Stanford University, Stanford, California 94305, USA}
\author{T.~Namikawa}
\affiliation{Kavli Institute for the Physics and Mathematics of the Universe (WPI), UTIAS, The University of Tokyo, Kashiwa, Chiba 277-8583, Japan}
\author{H.~T.~Nguyen}
\affiliation{Jet Propulsion Laboratory, Pasadena, California 91109, USA}
\author{R.~O'Brient}
\affiliation{Department of Physics, California Institute of Technology, Pasadena, California 91125, USA}
\affiliation{Jet Propulsion Laboratory, Pasadena, California 91109, USA}
\author{R.~W.~Ogburn~IV}
\affiliation{Department of Physics, Stanford University, Stanford, California 94305, USA}
\affiliation{Kavli Institute for Particle Astrophysics and Cosmology, SLAC National Accelerator Laboratory, 2575 Sand Hill Rd, Menlo Park, California 94025, USA}
\author{S.~Palladino}
\affiliation{Department of Physics, University of Cincinnati, Cincinnati, Ohio 45221, USA}
\author{T.~Prouve}
\affiliation{Service des Basses Temp\'{e}ratures, Commissariat \`{a} l'Energie Atomique, 38054 Grenoble, France}
\author{C.~Pryke}
\email{pryke@physics.umn.edu}
\affiliation{School of Physics and Astronomy, University of Minnesota, Minneapolis, Minnesota 55455, USA}
\affiliation{Minnesota Institute for Astrophysics, University of Minnesota, Minneapolis, Minnesota 55455, USA}
\author{B.~Racine}
\affiliation{Center for Astrophysics, Harvard \& Smithsonian, Cambridge, MA 02138, U.S.A}
\affiliation{Aix-Marseille  Universit\'{e},  CNRS/IN2P3,  CPPM,  13288 Marseille, France}
\author{C.~D.~Reintsema}
\affiliation{National Institute of Standards and Technology, Boulder, Colorado 80305, USA}
\author{S.~Richter}
\affiliation{Center for Astrophysics, Harvard \& Smithsonian, Cambridge, MA 02138, U.S.A}
\author{A.~Schillaci}
\affiliation{Department of Physics, California Institute of Technology, Pasadena, California 91125, USA}
\author{R.~Schwarz}
\affiliation{School of Physics and Astronomy, University of Minnesota, Minneapolis, Minnesota 55455, USA}
\author{B.~L.~Schmitt}
\affiliation{Center for Astrophysics, Harvard \& Smithsonian, Cambridge, MA 02138, U.S.A}
\author{C.~D.~Sheehy}
\affiliation{Physics Department, Brookhaven National Laboratory, Upton, NY 11973}
\author{A.~Soliman}
\affiliation{Department of Physics, California Institute of Technology, Pasadena, California 91125, USA}
\author{T.~St.~Germaine}
\affiliation{Center for Astrophysics, Harvard \& Smithsonian, Cambridge, MA 02138, U.S.A}
\affiliation{Department of Physics, Harvard University, Cambridge, MA 02138, USA}
\author{B.~Steinbach}
\affiliation{Department of Physics, California Institute of Technology, Pasadena, California 91125, USA}
\author{R.~V.~Sudiwala}
\affiliation{School of Physics and Astronomy, Cardiff University, Cardiff, CF24 3AA, United Kingdom}
\author{G.~P.~Teply}
\affiliation{Department of Physics, California Institute of Technology, Pasadena, California 91125, USA}
\author{K.~L.~Thompson}
\affiliation{Department of Physics, Stanford University, Stanford, California 94305, USA}
\affiliation{Kavli Institute for Particle Astrophysics and Cosmology, SLAC National Accelerator Laboratory, 2575 Sand Hill Rd, Menlo Park, California 94025, USA}
\author{J.~E.~Tolan}
\affiliation{Department of Physics, Stanford University, Stanford, California 94305, USA}
\author{C.~Tucker}
\affiliation{School of Physics and Astronomy, Cardiff University, Cardiff, CF24 3AA, United Kingdom}
\author{A.~D.~Turner}
\affiliation{Jet Propulsion Laboratory, Pasadena, California 91109, USA}
\author{C.~Umilt\`{a}}
\affiliation{Department of Physics, University of Cincinnati, Cincinnati, Ohio 45221, USA}
\affiliation{Department of Physics, University of Illinois at Urbana-Champaign, Urbana, Illinois 61801, USA}
\author{C.~Verg\`{e}s}
\affiliation{Center for Astrophysics, Harvard \& Smithsonian, Cambridge, MA 02138, U.S.A}
\author{A.~G.~Vieregg}
\affiliation{Department of Physics, Enrico Fermi Institute, University of Chicago, Chicago, IL 60637, USA}
\affiliation{Kavli Institute for Cosmological Physics, University of Chicago, Chicago, IL 60637, USA}
\author{A.~Wandui}
\affiliation{Department of Physics, California Institute of Technology, Pasadena, California 91125, USA}
\author{A.~C.~Weber}
\affiliation{Jet Propulsion Laboratory, Pasadena, California 91109, USA}
\author{D.~V.~Wiebe}
\affiliation{Department of Physics and Astronomy, University of British Columbia, Vancouver, British Columbia, V6T 1Z1, Canada}
\author{J.~Willmert}
\affiliation{School of Physics and Astronomy, University of Minnesota, Minneapolis, Minnesota 55455, USA}
\author{C.~L.~Wong}
\affiliation{Center for Astrophysics, Harvard \& Smithsonian, Cambridge, MA 02138, U.S.A}
\affiliation{Department of Physics, Harvard University, Cambridge, MA 02138, USA}
\author{W.~L.~K.~Wu}
\affiliation{Kavli Institute for Particle Astrophysics and Cosmology, SLAC National Accelerator Laboratory, 2575 Sand Hill Rd, Menlo Park, California 94025, USA}
\author{H.~Yang}
\affiliation{Department of Physics, Stanford University, Stanford, California 94305, USA}
\author{K.~W.~Yoon}
\affiliation{Department of Physics, Stanford University, Stanford, California 94305, USA}
\affiliation{Kavli Institute for Particle Astrophysics and Cosmology, SLAC National Accelerator Laboratory, 2575 Sand Hill Rd, Menlo Park, California 94025, USA}
\author{E.~Young}
\affiliation{Department of Physics, Stanford University, Stanford, California 94305, USA}
\affiliation{Kavli Institute for Particle Astrophysics and Cosmology, SLAC National Accelerator Laboratory, 2575 Sand Hill Rd, Menlo Park, California 94025, USA}
\author{C.~Yu}
\affiliation{Department of Physics, Stanford University, Stanford, California 94305, USA}
\author{L.~Zeng}
\affiliation{Center for Astrophysics, Harvard \& Smithsonian, Cambridge, MA 02138, U.S.A}
\author{C.~Zhang}
\affiliation{Department of Physics, California Institute of Technology, Pasadena, California 91125, USA}
\author{S.~Zhang}
\affiliation{Department of Physics, California Institute of Technology, Pasadena, California 91125, USA}
\date[Published in PRL 4 October 2021]{}

\begin{abstract}
We present results from an analysis of all data taken by the
\biceptwo, \keckarray\ and \bicepthree\ CMB polarization experiments
up to and including the 2018 observing season.
We add additional \keckarray\ observations at 220\,GHz
and \bicepthree\ observations at 95\,GHz to the previous 95/150/220\,GHz data set.
The $Q/U$ maps now reach depths of 2.8, 2.8 and 8.8\,\ukcmb\,arcmin at
95, 150 and 220\,GHz respectively over an effective area
of $\approx 600$ square degrees at 95\,GHz and $\approx 400$ square degrees
at 150 \& 220\,GHz.
The 220\,GHz maps now achieve a signal-to-noise on polarized dust emission
exceeding that of \planck\ at 353\,GHz.
We take auto- and cross-spectra between these maps and publicly
available \wmap\ and \planck\ maps at frequencies from 23 to 353\,GHz and
evaluate the joint likelihood of the spectra versus
a multicomponent model of lensed-\lcdm+$r$+dust+synchrotron+noise.
The foreground model has seven parameters, and no longer requires
a prior on the frequency spectral index of the dust emission taken
from measurements on other regions of the sky.
This model is an adequate description of the data at the current noise levels.
The likelihood analysis yields the constraint $r_{0.05}<\rul$ at
95\% confidence.
Running maximum likelihood search on simulations we
obtain unbiased results and find that $\sigma(r)=0.009$.
These are the strongest constraints to date on primordial gravitational waves.
\vspace{10mm} 
\end{abstract}


\maketitle

{\it Introduction.}---The \lcdm\ standard model of cosmology
is able to describe the observable universe in a statistical manner
using only six free parameters.
Measurements of the cosmic microwave background (CMB)~\cite{penzias65}
are one the key pillars of this model and now constrain
its parameters with percent-level precision
(see most recently Ref.~\cite{planck2018VI}).

The \lcdm\ model describes how the universe evolved
from an initial high energy state ($T\gg10^{12}$~K), and
the conditions at that time can be inferred from observations:
fractionally small, Gaussian, adiabatic perturbations
with a slightly red power law spectrum ($n_s \lesssim 1$).
Inflationary theories naturally explain such conditions as
the outcome of a pre-phase of exponential expansion during
which the scale of the proto-universe increased by a factor
of $\sim e^{60}$.
Inflation makes an additional prediction which has
not yet been observed---a background of tensor perturbations,
also known as gravitational waves
(see Ref.~\cite{kamionkowski2015} for a review and
citations to the original literature).
There are many specific inflationary models and classes
thereof.
If we can detect or set limits on primordial gravitational
waves we can set limits on these models~\cite{S4_sciencebook2016},
and probe physics at energy scales far higher than can ever
be accessed in laboratory experiments.

A polarization pattern can be decomposed into \emode\ (gradient)
and \bmode\ (curl) components.
Under the \lcdm\ standard model the CMB polarization pattern
is mostly \emode, with a much smaller \bmode\ component
which arises due to gravitational deflections (lensing) of the CMB
photons after their last scattering~\cite{zaldarriaga98}.
Since primordial gravitational waves will produce
{\emode}s and {\bmode}s approximately equally
it was realized in the late 1990's that the best way to search
for them is to look for an excess {\bmode}
signal~\cite{seljak97b,kamionkowski97,seljak97a}.
Additional non-primordial {\bmode}s are produced by astrophysical
foreground emissions, primarily from our own galaxy, but these
have different frequency spectra than the CMB, and can
be separated from it using multi frequency measurements.

Our \bicep/\keck\ program first reported detection of an
excess over the lensing \bmode\ expectation at
150\,GHz in Ref.~\cite{biceptwoI}.
In a joint analysis using multi-frequency data from
the \planck\ experiment it was shown that most
or all of this is due to polarized emission from dust in our own
galaxy~\cite[hereafter BKP]{bkp}.
In Ref.~\cite[hereafter BK14]{biceptwoVI} we improved the
constraint using \keckarray\ data at 95\,GHz taken during
the 2014 season, and in Ref.~\cite[hereafter BK15]{biceptwoX}
we improved again adding \keckarray\ data at 95\,GHz and 220\,GHz
taken during the 2015 season.
In this letter [hereafter BK18] we add large amounts of new
data taken by \keckarray\ at 220\,GHz and \bicepthree\
at 95\,GHz during the 2016, 2017 and 2018 observing seasons.
This paper follows BK15 very closely in the methods, structure,
and, in places, even the wording, mainly just adding additional experimental data.
This improves the constraint on primordial gravitational waves parameterized by
the tensor-to-scalar ratio $r$ by more than a factor of two over our
previous result to $r_{0.05}<\rul$ at 95\% confidence,
setting important additional limits on inflationary models.

{\it Instrument and observations.}---The \biceptwo\ receiver
observed at 150\,GHz from 2010--2012~\cite{biceptwoII}.
The \keckarray\ was essentially five copies of \biceptwo\ running
in parallel from 2012--2019, initially at 150\,GHz but
switching over time to 95 and 220\,GHz~\cite{biceptwoV}.
\bicepthree\ is a single similar, but scaled up, receiver which commenced
science observations in the 2016 Austral winter season~\cite{biceptwoXV}.
Whereas the \biceptwo\ and \keck\ 150 \& 220\,GHz receivers each
contained $\approx 500$ bolometric detectors \bicepthree\ contains
$\approx 2500$ detectors.
The aperture size is also increased from $\approx 0.25$\,m to $\approx 0.5$\,m.
The \keck\ receivers were mounted on a single telescope mount (movable platform),
while \bicepthree\ occupies a separate mount previously used
for \biceptwo\ on a nearby building.
All of these telescopes are located at the South Pole Station in Antarctica.
The mounts scan the receivers across the sky,
and the cryogenic detectors track the intensity of the incoming microwave radiation.
The detectors are arranged as interleaved orthogonally polarized pairs
in the focal planes and the pair difference timestreams are thus
measures of the polarized emission from the sky~\cite{bkdets}.
At the South Pole the atmosphere is exceptionally transparent
and stable at the observation frequencies~\cite[Fig.~5]{kuo17}.

\biceptwo\ and \keckarray\ both mapped a region of sky centered at
RA 0h, Dec.\ $-57.5\deg$ with an effective area of $\approx 400$ square degrees.
\bicepthree\ has a larger instantaneous field of view and hence naturally maps
a larger sky area with an effective area of $\approx 600$ square degrees.
We have perturbed the center of the \bicepthree\ scan region such that
most of this additional area falls on the higher declination side
of the sky patch in an attempt to stay away from regions where the \planck\
data indicates polarized dust contamination may be higher.
The BK15 data set consisted of 4/17/2 receiver-years
at 95/150/220\,GHz respectively.
\bicepthree\ is equivalent to about eight of the \keckarray\ 95\,GHz
receivers~\cite{biceptwoXV}
so the BK18 data set is equivalent to about 28/18/14 \keck\ receiver-years
at 95/150/220\,GHz respectively.

{\it Maps and Power Spectra}---We make maps and power spectra
using the same procedures as in our previous series of papers.
The timestream data are binned into pixels on the sky using
knowledge of the pointing direction of the telescope
at each moment in time, together with the relative angles
from the telescope boresight to each individual detector
pair.
By taking data with the receivers rotated at a range
of angles, maps of the Stokes parameters $Q$ and $U$ can be constructed.

The maps at each observing frequency are subjected to
a matrix purification operation~\cite{biceptwoI,biceptwoVII} such that they contain
only structures sourced by {\bmode}s of the underlying
sky pattern.
This allows us to measure the {\bmode}s in the presence of the
much brighter \lcdm\ {\emode}s.
The maps are then inverse noise variance apodized, Fourier
transformed and rotated from the $Q/U$ to the $E/B$ basis.
In this paper we use our own maps at 95, 150 and 220\,GHz
plus the 23 \& 33\,GHz bands of
\wmap~\footnote{See \url{http://lambda.gsfc.nasa.gov/product/map/dr5/m_products.cfm}}\citep{bennett13}
and the 30, 44, 143, 217 and 353\,GHz maps from the NPIPE processing of the \planck\
data~\footnote{See \url{https://irsa.ipac.caltech.edu/data/Planck/release_3/ancillary-data/HFI_Products.html}}\citep{planckiLVII}.
For illustration purposes we can inverse Fourier transform to
form $E/B$ maps.
Fig.~\ref{fig:eb_maps} shows $E$- and \bmode\ maps at
95, 150 and 220\,GHz.
(See Appendix~\ref{app:maps} for the full set of $T/Q/U$ maps.)

\begin{figure*}
\resizebox{\textwidth}{!}{\includegraphics{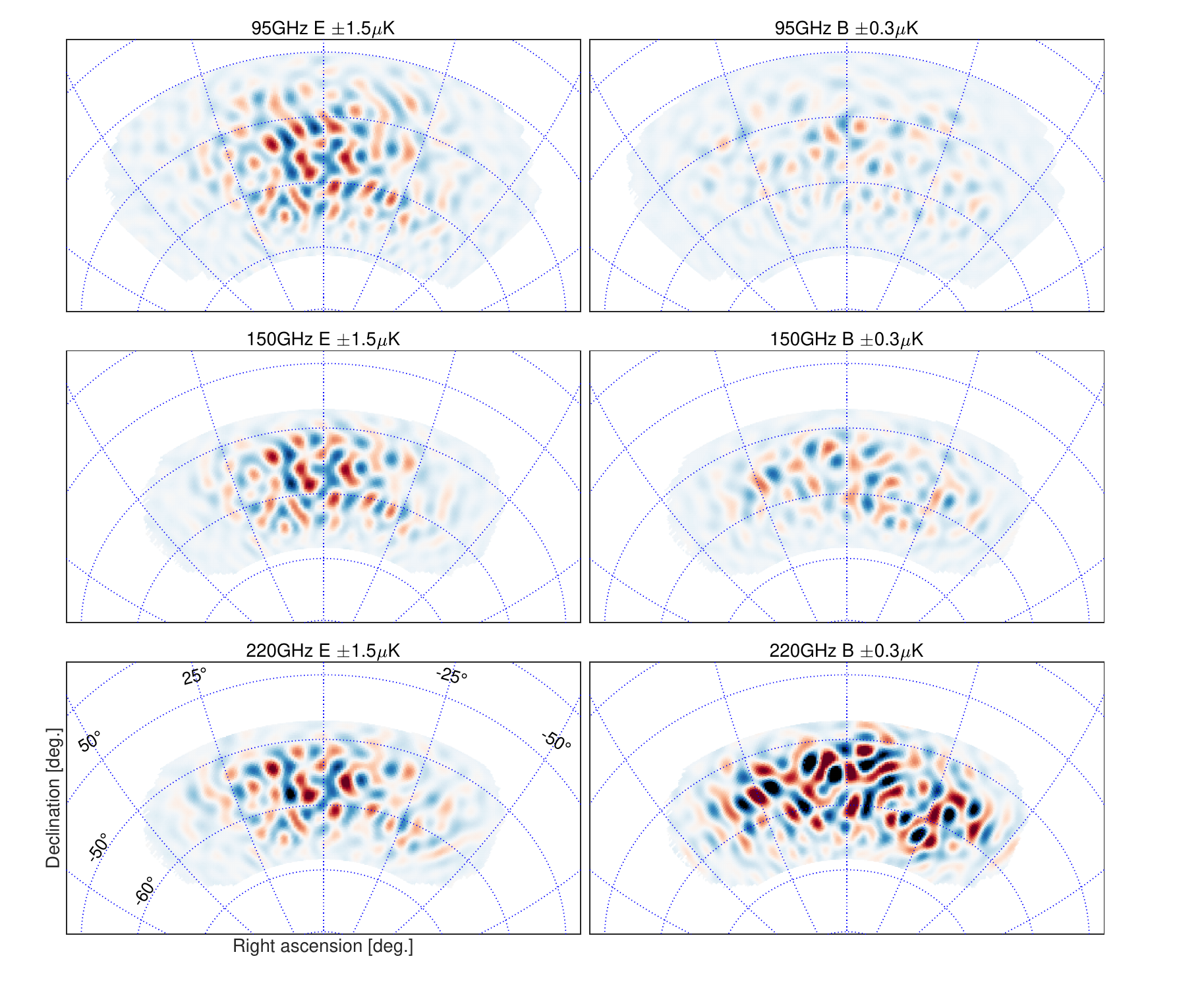}}
\caption{\emode\ (left column) and \bmode\ (right column) maps at
95, 150 and 220\,GHz in CMB units, and filtered to degree
angular scales ($50<\ell<120$).
Note the differing color ranges left and right.
The $E$ maps are dominated by \lcdm\ signal, and hence
are highly correlated across all three bands.
The 95\,GHz $B$ map is approximately equal parts lensed-\lcdm\
signal and noise.
At 150 and 220\,GHz the $B$ maps are dominated by polarized
dust emission.}
\label{fig:eb_maps}
\end{figure*}

\begin{figure*}
\resizebox{0.8\textwidth}{!}{\includegraphics{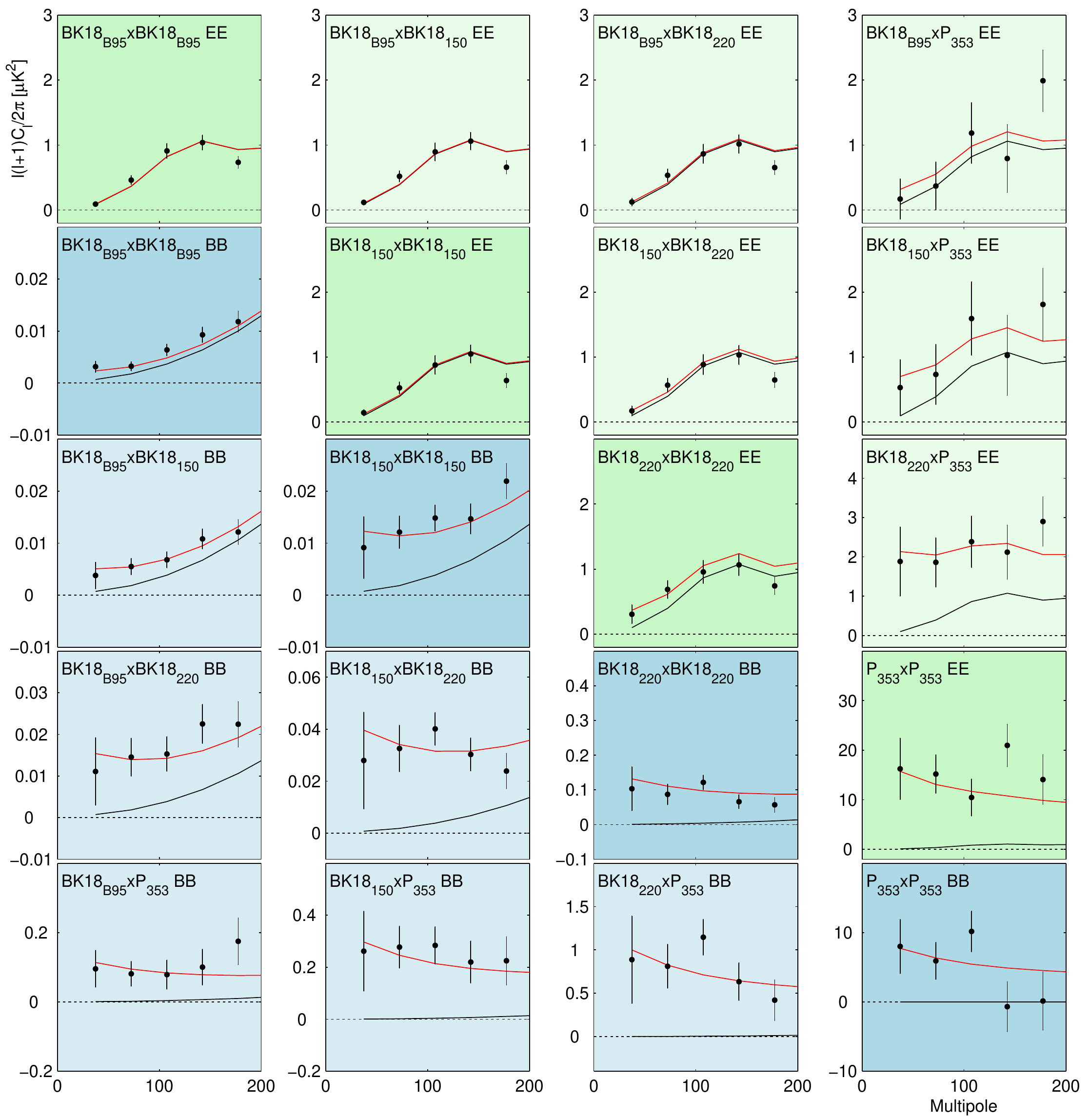}}
\caption{$EE$ (green) and $BB$ (blue) auto- and cross-spectra calculated using
the \bicepthree\ 95\,GHz map, the \biceptwo/\keck\ 150\,GHz map, the
\keck\ 220\,GHz map, and the \planck\ 353\,GHz map (with the auto-spectra
in darker colors).
The \bicep/\keck\ maps use all data taken up to and including
the 2018 observing season---we refer to these as BK18.
The black lines show the model expectation values
for lensed-\lcdm, while the red lines show the expectation values of a
baseline lensed-\lcdm+dust model from our previous BK15 analysis
($r=0$, $\Adf=4.7$\,\uksq, $\Bd=1.6$, $\ad=-0.4$).
Note that the model shown was fit to $BB$ only and did not
use the \bicepthree\ 95\,GHz points shown (which are entirely new).
The agreement with the spectra involving 95\,GHz and all the
$EE$ spectra (under the assumption that $EE/BB=2$ for dust)
is therefore a validation of the model.
}
\label{fig:powspecres_bkbands}
\end{figure*}

We take the variance within annuli of the Fourier plane
to estimate the angular power spectra.
Fig.~\ref{fig:powspecres_bkbands} shows the $EE$ and $BB$
auto- and cross-spectra for the \bicep/\keck\ bands plus the
\planck\ 353~GHz band which remains important for constraining the
polarized dust contribution.
Comparing this plot to Fig.~2 of BK15 we can see that the uncertainties
are dramatically reduced for the auto- and cross-spectra of
the 95 and 220\,GHz bands.
The model plotted is a ``baseline'' lensed-\lcdm+dust model
from our previous BK15 analysis, which remains a good
description of the data.
The $EE$ spectra were not used to derive the model but
agree well with it under the assumption that $EE/BB=2$ for
dust, as is known to be close to the case~\cite{planckiXXX,planck2018XI}.

\begin{figure}
\resizebox{\columnwidth}{!}{\includegraphics{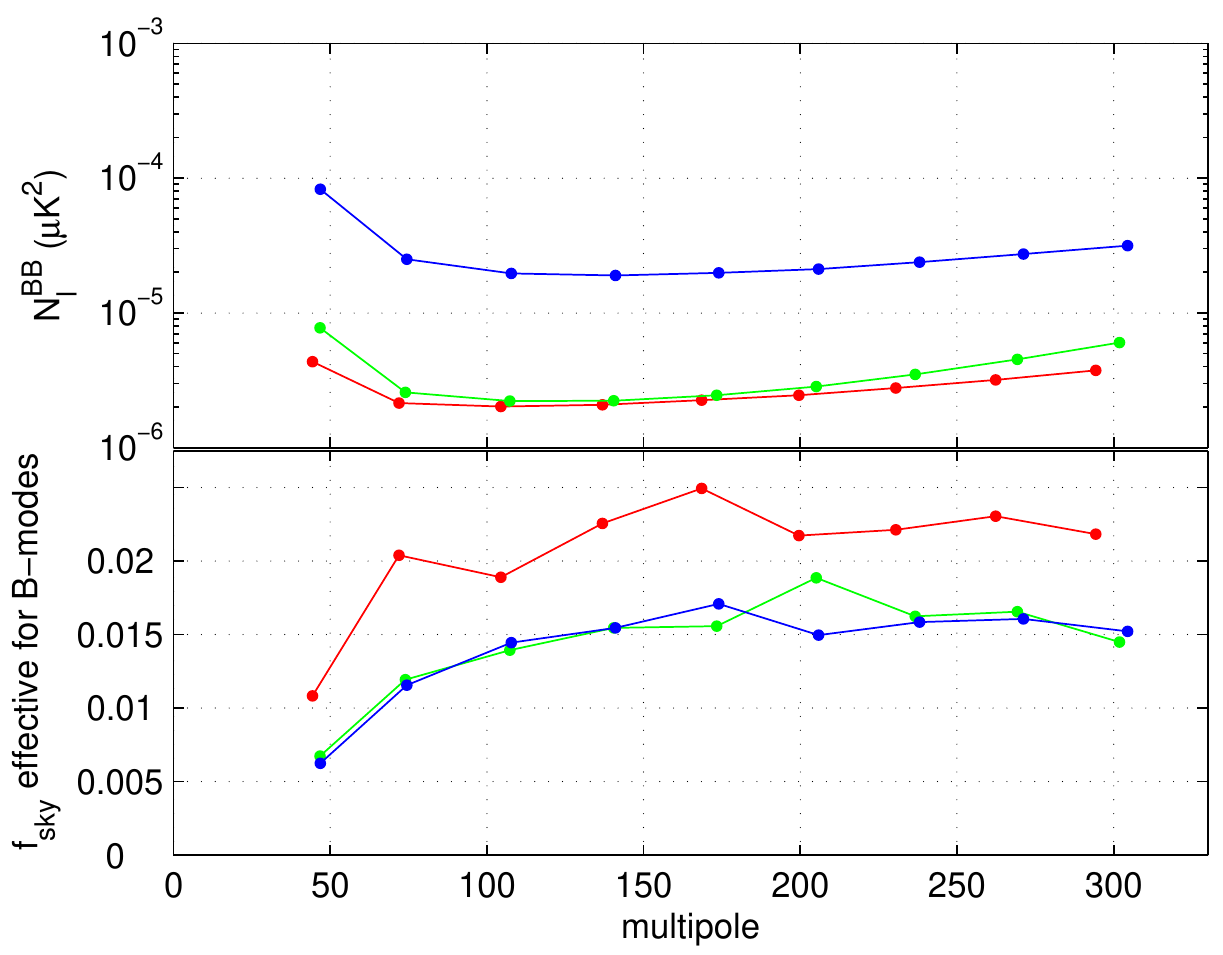}}
\caption{{\it Upper:} The noise spectra of the \bicepthree\
95\,GHz map (red), the \biceptwo/\keck\ 150\,GHz map
(green) and the \keck\ 220\,GHz maps (blue).
The spectra are shown after correction for the filtering of signal which occurs
due to the beam roll-off, timestream filtering, and \bmode\ purification.
(Note that no $\ell^2$ scaling is applied.)
{\it Lower:} The effective sky fraction as calculated from
the ratio of the mean noise realization bandpowers to their fluctuation
$f_\mathrm{sky}(\ell)=\frac{1}{2\ell \Delta \ell} \left( \frac{\sqrt{2}\bar{N_b}}{\sigma(N_b)} \right)^2$,
i.e.\ the observed number of {\bmode} degrees of freedom divided by the
nominal full-sky number.
The turn-down at low $\ell$ is due to mode loss to the timestream filtering and
matrix purification.
}
\label{fig:nl_fsky}
\end{figure}

To test for systematic contamination we carry out our usual
``jackknife'' internal consistency (null) tests on the new 95\,GHz and 220\,GHz
data as described in Appendices~\ref{app:mapjack}
and~\ref{app:specjack}.
Fig.~\ref{fig:nl_fsky} upper shows the noise spectra
for the three main BK18 bands after correction for the filter and beam
suppression.
In an auto-spectrum the quantity which determines the ability to constrain 
$r$ is the fluctuation of the noise bandpowers rather than their mean.
The lower panel therefore shows the effective sky fraction
as inferred from the fractional noise fluctuation.
Together, these panels provide a useful synoptic measure of the
loss of information due to noise, filtering, and $EE/BB$
separation in the lowest bandpowers (and we are glad to see
taken up by others as e.g. Fig.~6 of Ref.~\cite{polarbear19}).

{\it Likelihood Analysis.}---We perform likelihood analysis using
the methods introduced in BKP and refined in BK14 \& BK15.
We use the Hamimeche-Lewis approximation~\citep{hamimeche08} to the joint
likelihood of the ensemble of 66 $BB$ auto- and cross-spectra taken
between the \bicep/\keck, \wmap\ and \planck\ maps.
We compare the observed bandpower values for ${20<\ell<330}$
(9 bandpowers per spectrum)
to an eight parameter model of lensed-\lcdm+$r$+dust+synchrotron+noise
and explore the parameter space using \texttt{COSMOMC}~\citep{cosmomc}
(which implements a Markov chain Monte Carlo method).
As in our previous analyses the bandpower covariance matrix is derived
from 499 simulations of signal and noise, explicitly setting to zero terms
such as the covariance of signal-only bandpowers with noise-only bandpowers
or covariance of \bicep/\keck\ noise bandpowers with \wmap/\planck\ noise bandpowers
(see Appendix H of BK15 and Appendix B of Ref.~\cite{s4forecast20} for
details).
We deal with the differing sky coverage of the \bicepthree\ and
\biceptwo/\keck\ maps as described in Appendix~\ref{app:allspec}.
The tensor/scalar power ratio $r$ is evaluated at a pivot scale of
0.05~Mpc$^{-1}$, and we fix the tensor spectral index $n_t=0$.
A \texttt{COSMOMC} module containing the data and model
is available for download at \url{http://bicepkeck.org}.
The following paragraphs briefly summarize the foreground model.

We include dust with amplitude $\Adf$ evaluated
at 353\,GHz and ${\ell=80}$.
The frequency spectral behavior is taken as a modified black body spectrum
with ${T_\mathrm{d}=19.6}$\,K and frequency spectral index $\Bd$.
In a significant change from the baseline analysis choices of BK15,
we remove the prior on the dust frequency spectral index
which was previously applied based on \planck\ data in other
regions of sky---with the improvement
in the \keck\ 220\,GHz sensitivity this prior is no longer needed.
The spatial power spectrum is taken as a power law
${\mathcal{D}_\ell \propto \ell^{\ad}}$ marginalizing
uniformly over the (generous) range ${-1<\ad<0}$
(where ${\mathcal{D}_\ell \equiv \clstar}$).
\planck\ analysis consistently finds approximate power
law behavior of both the $EE$ and $BB$ dust spectra with
exponents $\approx -0.4$ ~\cite{planckiXXX,planck2018XI}.

We include synchrotron with amplitude $\Asf$
evaluated at 23\,GHz (the lowest \wmap\ band) and $\ell=80$,
assuming a simple power law for the frequency spectral behavior
${\As \propto \nu^{\Bs}}$, and using a Gaussian prior
${\Bs=-3.1\pm0.3}$ taken from the analysis of \wmap\ 23 and 33\,GHz data
in Ref~~\citep{fuskeland14}.
We note that analysis of 2.3\,GHz data from \spass\ in
conjunction with \wmap\ and \planck\ finds $\Bs=-3.2$ with no detected
trends with galactic latitude or angular scale~\cite{krachmalnicoff18},
and that Ref.~\cite{Fuskeland2020} analyzed the \spass\ and
\wmap\ 23\,GHz data and found $\Bs=-3.22\pm0.06$ in the \biceptwo\
sky patch.
The spatial power spectrum is taken as a power law
${\mathcal{D}_\ell \propto \ell^{\as}}$ marginalizing
over the range ${-1<\as<0}$~\cite{dunkley08}.
Ref.~\cite{krachmalnicoff18} finds a value at the
bottom end of this range ($\approx -1$) from the \spass\ data
for $BB$ at high galactic latitude.

Finally we include sync/dust correlation parameter $\epsilon$
(called $\rho$ in some other papers~\cite{choi15,planck2018XI,krachmalnicoff18}).
As in BK15 we marginalize over the full possible range
${-1<\epsilon<1}$.

\begin{figure*}
\begin{center}
\resizebox{0.7\textwidth}{!}{\includegraphics{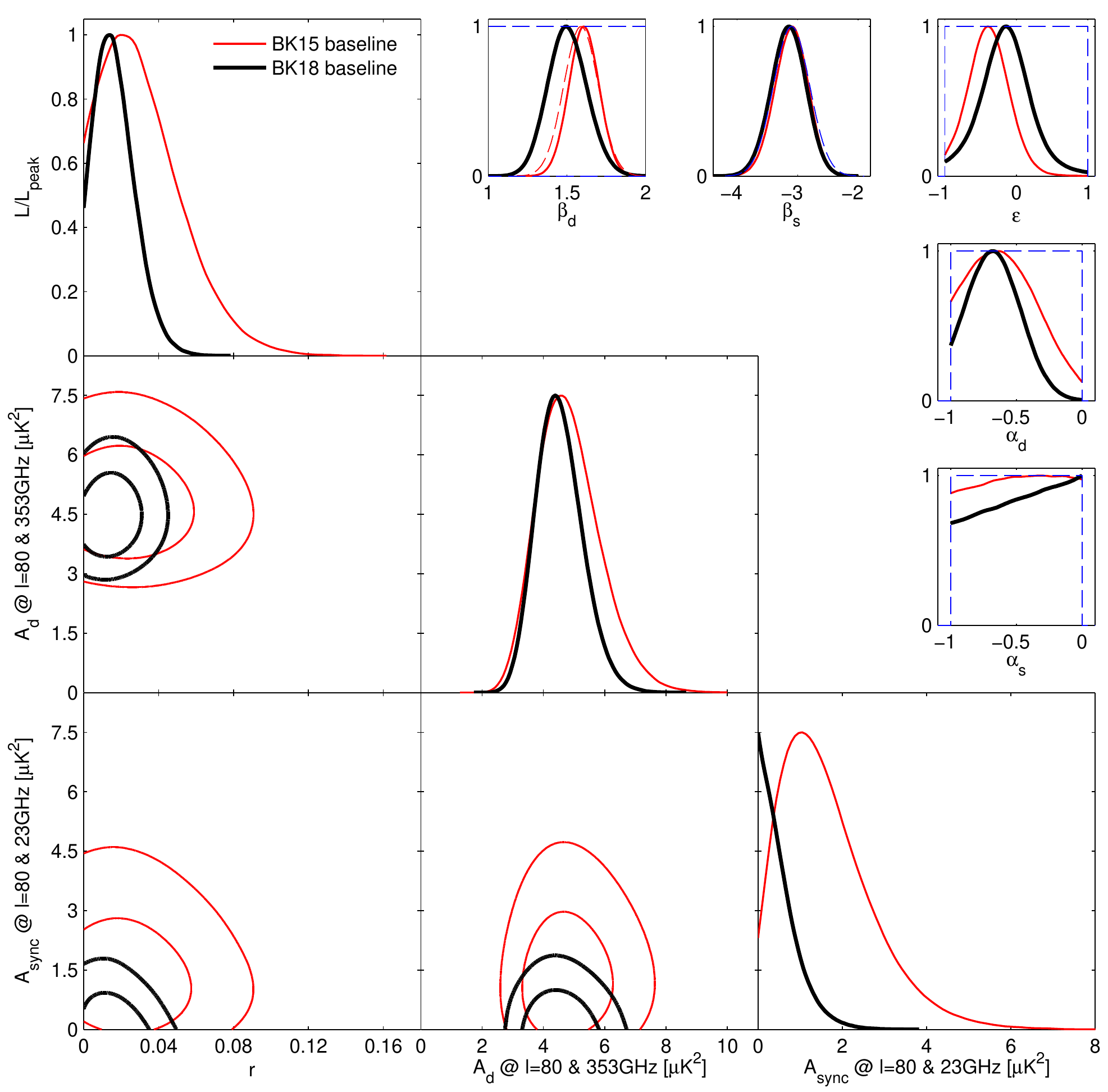}}
\end{center}
\caption{Results of a multicomponent multi-spectral likelihood
analysis of \bicep/\keck+\wmap/\planck\ data.
The red faint curves are the baseline result from the previous
BK15 paper (the black curves from Fig.~4 of that paper).
The bold black curves are the new baseline BK18 results,
adding a large amount of additional data at 95 and 220\,GHz taken by
\bicepthree\ and \keckarray\ during the 2016--2018 observing seasons.
The upper limit on the tensor-to-scalar ratio tightens
to $r_{0.05}<\rul$ at 95\protect\% confidence.
The parameters $\Ad$ and $\As$ are the
amplitudes of the dust and synchrotron $B$-mode power spectra, where
$\beta$ and $\alpha$ are the respective frequency and spatial
spectral indices.
The correlation coefficient between the dust and synchrotron patterns is
$\epsilon$.
In the $\beta$, $\alpha$ and $\epsilon$ panels the dashed lines
show the priors placed on these parameters (either Gaussian or uniform).
Note that the Gaussian prior on $\Bd$ has been removed going from BK15 to BK18.
}
\label{fig:likebase}
\end{figure*}

We hold the lensing \bmode\ spectrum fixed at that predicted for
the \planck\ 2018 cosmological parameters~\cite[Table 2]{planck2018VI}.
Results of our baseline analysis are shown in
Fig.~\ref{fig:likebase} and yield the following
statistics:
$r_{0.05}=\rrange$ ($r_{0.05}<\rul$ at 95\% confidence),
$\Adf=\Adrange$\,\uksq,
$\Asf<\Asul$\,\uksq\ at 95\% confidence,
and $\Bd=\Bdrange$.
For $r$, the zero-to-peak likelihood ratio is \rztop.
Taking
${\frac{1}{2} \left( 1-f \left( -2\log{L_0/L_{\rm peak}} \right) \right)}$,
where $f$ is the $\chi^2$ CDF (for one degree of freedom),
we estimate that the probability to get a likelihood ratio smaller than this is
\rztopps\% if, in fact, $r=0$.
As compared to the previous BK15 analysis, the likelihood curve
for $r$ tightens considerably with the peak position shifting down slightly, and
the $\Ad$ curve tightens slightly.
In addition the $\As$ curve now peaks at zero---the weak evidence for
synchrotron we saw in BK15 is no longer present.
(Using the \spass\ data~\cite{krachmalnicoff18} we estimate that the expectation
is $\Asf\approx0.4$\,\uksq\ in the \bicep/\keck\ field, which is consistent with
our $\As$ likelihood curve.)
In the BK15 analysis the constraint on $\Bd$ was prior dominated, but
for BK18 we see that the data is able to constrain this parameter
almost as well as the prior previously did.
Interestingly the peak value selected is very close to the mean value
from \planck\ 2018 analysis of larger regions of sky
${\Bd=1.53}$~\cite{planck2018XI}.

The maximum likelihood model has parameters
$r_{0.05}=\rmlm$, $\Adf=\Admlm$\,\uksq,
$\Asf=\Asmlm$\,\uksq,
$\Bd=\Bdmlm$, $\Bs=\Bsmlm$,
$\ad=\admlm$, $\as=\asmlm$,
and $\epsilon=\emlm$.
This model is an acceptable fit to the data with the probability to exceed (PTE)
the observed value of $\chi^2$ being \chitwoptesim.
Thus, while the dust spectrum might in general be expected
to exhibit fluctuations about power law spatial spectral behavior
greater than that expected for a Gaussian random field, for the
present the model continues to be an adequate description of the
data---see Appendix~\ref{app:allspec} for further details.

In Appendix~\ref{app:likeevolvar} we explore variation
and validation of the likelihood.
In Appendix~\ref{app:likevar} we vary the baseline analysis choices
and data selection, finding that these do not significantly alter
the results, and that the data do not prefer allowing decorrelation
of the dust pattern in the model.
We also find that the value of $\Ad$ is very similar when evaluated
over the larger \bicepthree\ sky coverage region and the smaller
\biceptwo/\keck\ sky region.
Freeing the amplitude of the lensing power we obtain
$A_{\rm L}^{\rm BB}= 1.03^{+0.08}_{-0.09}$, and the $r$ constraint hardly changes.
In Appendix~\ref{app:likevalid} we verify that the likelihood analysis is unbiased,
and in Appendix~\ref{app:altfore} we explore a suite of alternate
foreground models.
As part of our standard data reduction we ``deproject'' leading order
temperature to polarization leakage~\cite{biceptwoI,biceptwoIII}---in
Appendix~\ref{app:syst} we quantify possible residual leakage
and some other possible systematics.

\begin{figure}
\begin{center}
\resizebox{\columnwidth}{!}{\includegraphics{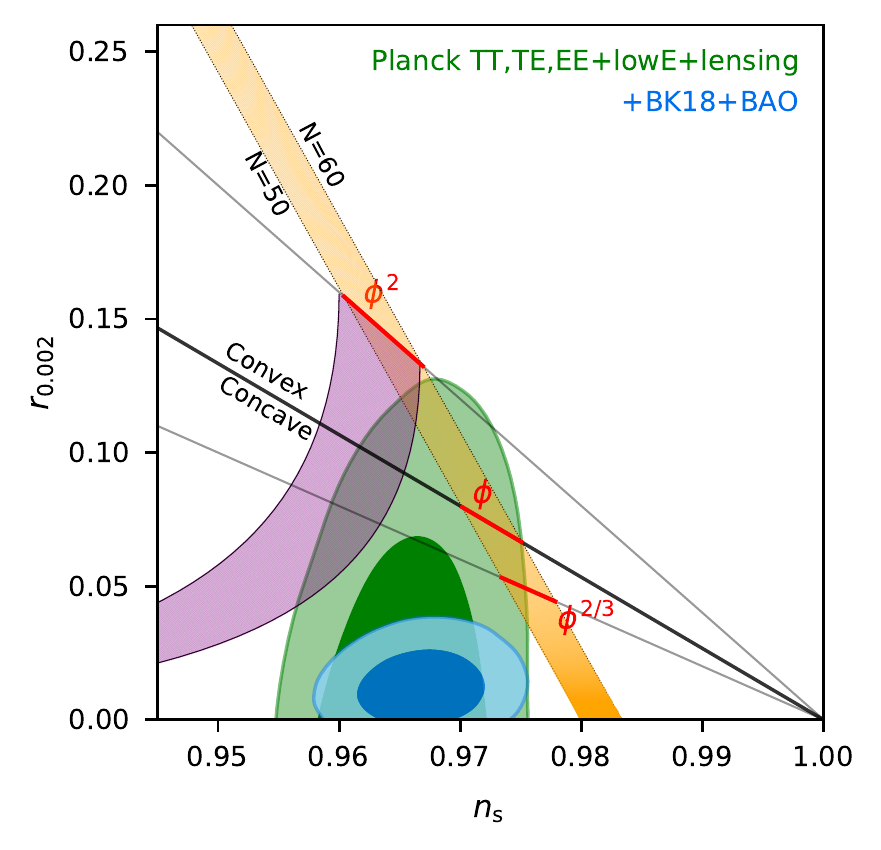}}
\end{center}
\caption{
Constraints in the $r$ vs.\ $n_s$ plane for the \planck\ 2018
baseline analysis,
and when also adding \bicep/\keck\ data through the end of the 2018
season plus BAO data to improve the constraint on $n_s$.
The constraint on $r$ tightens from $r_{0.05}<0.11$ to $r_{0.05}<0.035$.
This figure is adapted from Fig.~28 of Ref.~\cite{planck2018VI}
with the green contours being identical.
Some additional inflationary models are added from Fig.~8 of
Ref.~\cite{planck2018X} with the purple region being
natural inflation.
}
\label{fig:rns}
\end{figure}

Fig.~\ref{fig:rns} shows the constraints in the
$r$ vs. $n_s$ plane for the \planck\ 2018 baseline analysis~\cite{planck2018VI}
and when adding in BK18 \& BAO.
The BK18 data shrinks the contours in the vertical ($r$) direction while
the BAO data shrinks the contours in the horizontal ($n_s$) direction and
shifts the centroid slightly to the right.
The $\phi^{2/3}$ model now lies outside the 95\% contour
as does the band of natural inflation models.

{\it Conclusions.}---The BKP analysis yielded a 95\% confidence
constraint $r_{0.05}<0.12$, which BK14 improved to
$r_{0.05}<0.09$, and BK15 improved to $r_{0.05}<0.07$.
The BK18 result described in this letter, $r_{0.05}<\rul$,
represents a fractional improvement equivalent to the two
previous steps combined.
The BK18 simulations have a median 95\% upper limit
of $r_{0.05}<0.019$. 

The distributions of maximum likelihood $r$ values
in simulations where the true value of $r$ is zero gave
$\sigma(r_{0.05})=0.020$ for BK15 which is reduced to $\sigma(r_{0.05})=0.009$
for BK18 (see Appendix~\ref{app:likevalid} for details).
Such simulations can also be used to investigate the degree to
which the analysis is limited by foregrounds and lensing.
Running the baseline BK18 analysis on simulations which contain no lensing
{\bmode}s gives $\sigma(r_{0.05})=0.004$, while running without
foreground parameters on simulations which contain no dust gives
$\sigma(r_{0.05})=0.007$.
Running without foreground parameters on simulations which
contain neither lensing or dust gives $\sigma(r_{0.05})=0.002$.

\begin{figure}
\resizebox{\columnwidth}{!}{\includegraphics{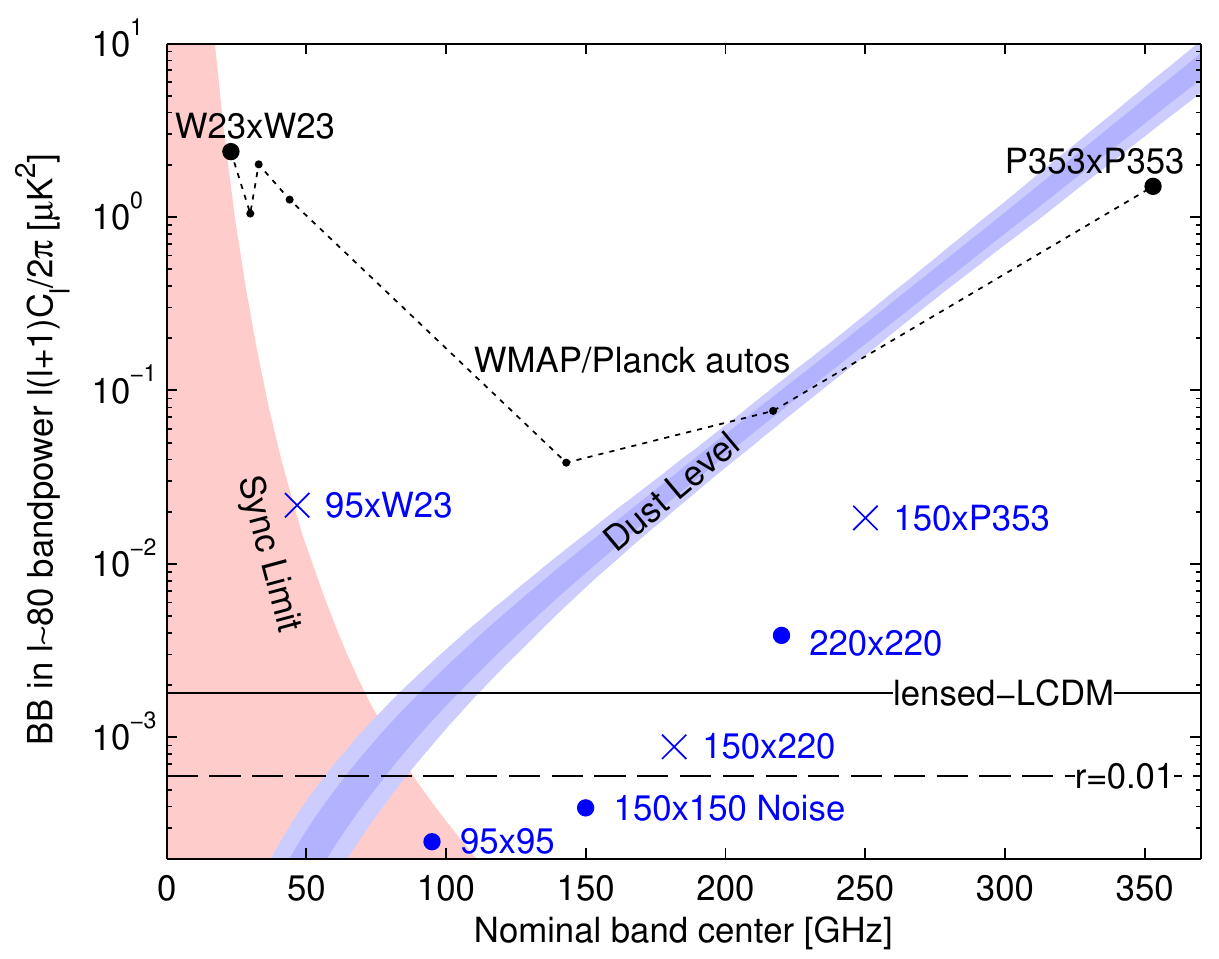}}
\caption{
Expectation values and noise uncertainties
of the $\ell\sim80$ $BB$ bandpower in the \bicep/\keck\ field.
The solid and dashed black lines show the expected signal power of
lensed-\lcdm\ and $r_{0.05}=0.01$.
Since CMB units are used, the levels corresponding
to these are flat with frequency.
The blue bands show the 1 and $2\sigma$ ranges of dust, and
the red shaded region shows the $95\%$ upper limit on
synchrotron in the baseline analysis including the uncertainties
in the amplitude and frequency spectral index parameters
($\Asf,\Bs$ and $\Adf,\Bd$).
The \bicep/\keck\ auto-spectrum noise uncertainties are shown as large blue
circles, and the noise uncertainties of the used \wmap/\planck\ single-frequency
spectra evaluated in the \bicep/\keck\ field are shown in black.
The blue crosses show the noise uncertainty of selected cross-spectra,
and are plotted at horizontal positions such that they
can be compared vertically with the dust and sync curves.}
\label{fig:noilev}
\end{figure}

Fig.~\ref{fig:noilev} shows the BK18 noise uncertainties in the $\ell\approx 80$
bandpowers as compared to the signal levels.
The signal-to-noise on polarized dust emission of our 220\,GHz band
is now considerably higher than that of the \planck\ 353\,GHz band---i.e.\
the $220\times220$ noise point is much further below the dust band
than the P353$\times$P353 point.
Additional \bicepthree\ data taken during 2019--2021 will reduce
the noise by a factor greater than 2 and $\sqrt{2}$ for
$95\times95$ and $95\times$W23 respectively, and we
have also recorded additional data at 220 and 270\,GHz.

\begin{figure}
\begin{center}
\resizebox{\columnwidth}{!}{\includegraphics{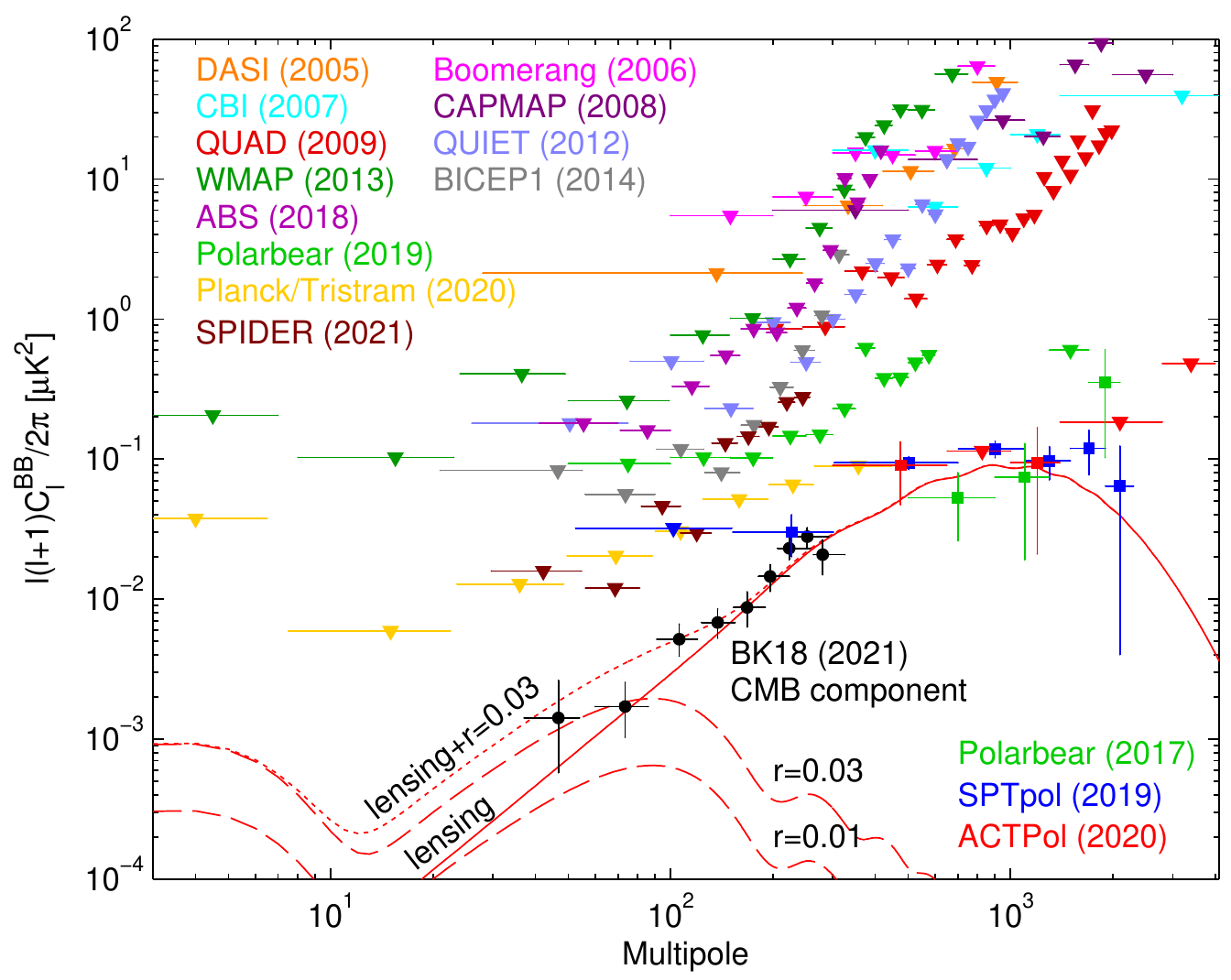}}
\end{center}
\caption{Summary of CMB \bmode\ polarization upper
limits~\cite{leitch05,montroy06,sievers07,bischoff08,brown09,quiet12,bennett13,barkats14,abs18,polarbear19,tristram20,spider21}
and detections~\cite{sptpol19,actpol20,polarbear17}.
Theoretical predictions are shown for the lensing {\bmode}s (solid red) which peak
at arcminute scales (multipole $\ell \sim 1000$), and for
gravitational wave {\bmode}s (dashed red) for two values of $r$
peaking at degree scales ($\ell \sim 80$).
The BK18 data are shown after removing Galactic foregrounds.}
\label{fig:bk_vs_world}
\end{figure}

Fig.~\ref{fig:bk_vs_world} shows the estimated CMB-only component of the BK18
\bmode\ bandpowers versus measurements from other experiments.
See Appendix~\ref{app:allspec} for a description of how the CMB-only power spectrum
estimate is calculated.

Figure~\ref{fig:powspecres_bkbands} shows that the BK18 data
is consistent with \lcdm\ plus a remarkably simple dust only
foreground model.
Nevertheless as we move forward to even higher levels of sensitivity
dust decorrelation, and foreground complexity more generally,
will remain a serious concern.
In addition, we are already in the regime where the sample
variance of the lensing component dominates $\sigma(r)$.
However, the lensing {\bmode}s can be spatially
separated from a primordial component and in this regard
we have recently demonstrated a path forward
by adding a ``lensing template'' derived from \sptpol\ and
\planck\ data to the BK14 analysis, resulting in
an improved constraint on $r$~\cite{bk14sptpol}.

The \keckarray\ mount has now been replaced by a larger,
more capable machine and we are in the process of upgrading
to a new system we call \biceparray~\cite{bkspie18ba,bkspie20rx}.
A \bicepthree\ class receiver is now operating in the 30/40\,GHz
band and in the coming years additional receivers will be
installed at 95, 150 and 220/270\,GHz.
The system is projected to reach $\sigma(r)\sim0.003$ within five years
with delensing in conjunction with \spt3G.

\acknowledgments

The \bicep/\keck\ projects have been made possible through
a series of grants from the National Science Foundation
including 0742818, 0742592, 1044978, 1110087, 1145172, 1145143, 1145248,
1639040, 1638957, 1638978, \& 1638970, and by the Keck Foundation.
The development of antenna-coupled detector technology was supported
by the JPL Research and Technology Development Fund, and by NASA Grants
06-ARPA206-0040, 10-SAT10-0017, 12-SAT12-0031, 14-SAT14-0009
\& 16-SAT-16-0002.
The development and testing of focal planes were supported
by the Gordon and Betty Moore Foundation at Caltech.
Readout electronics were supported by a Canada Foundation
for Innovation grant to UBC.
Support for quasi-optical filtering was provided by UK STFC grant ST/N000706/1.
The computations in this paper were run on the Odyssey/Cannon cluster
supported by the FAS Science Division Research Computing Group at
Harvard University.
The analysis effort at Stanford and SLAC is partially supported by
the U.S. DOE Office of Science.
We thank the staff of the U.S. Antarctic Program and in particular
the South Pole Station without whose help this research would not
have been possible.
Most special thanks go to our heroic winter-overs Robert Schwarz,
Steffen Richter, Sam Harrison, Grantland Hall and Hans Boenish.
We thank all those who have contributed past efforts to the \bicep/\keck\
series of experiments, including the \bicepone\ team.
We also thank the \planck\ and \wmap\ teams for the use of their
data, and are grateful to the \planck\ team for helpful discussions.
\vspace{100pt} 

\bibliography{ms}

\clearpage

\begin{appendix}

\section{Maps}
\label{app:maps}

\begin{figure*}
\resizebox{\textwidth}{!}{\includegraphics{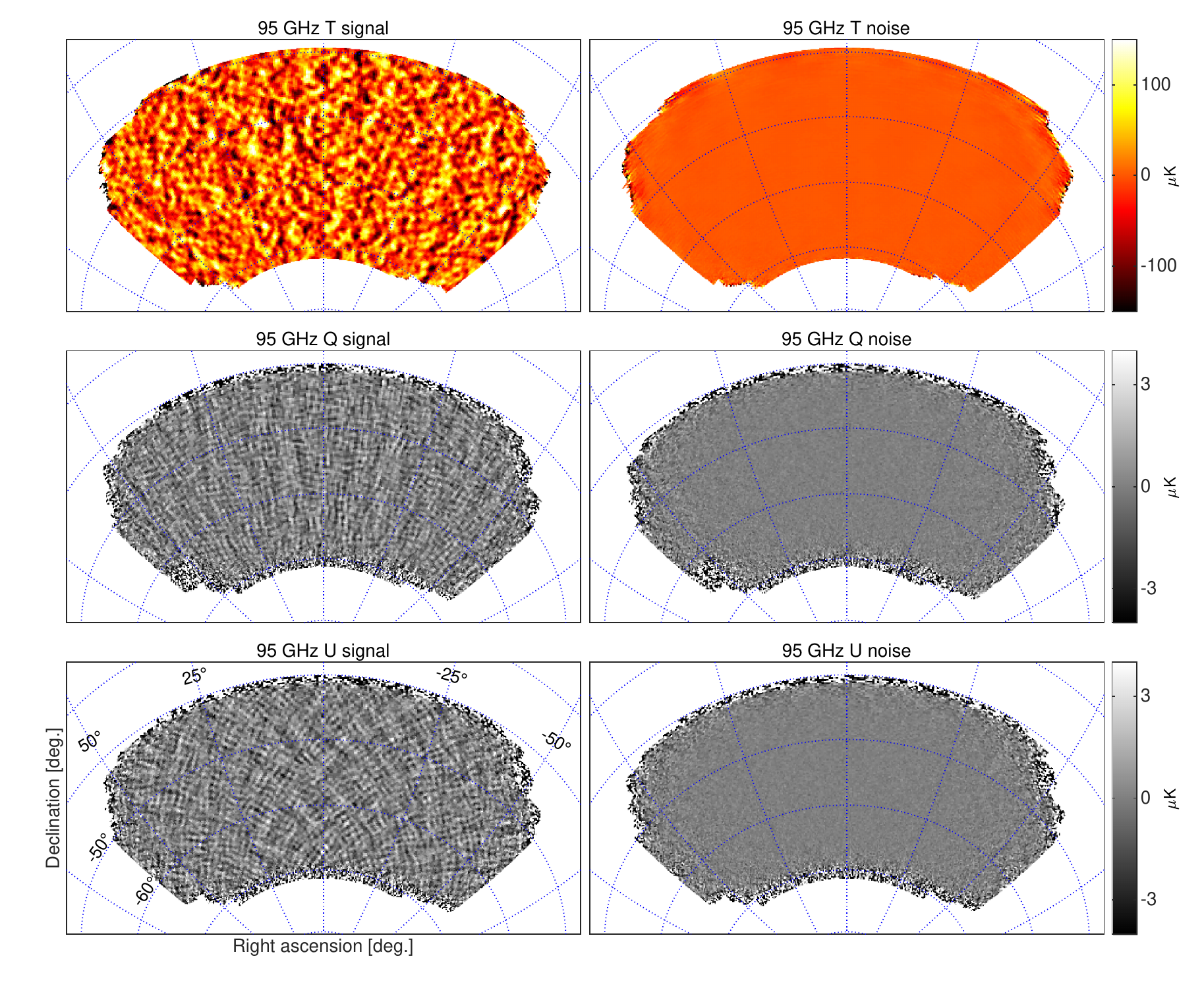}}
\caption{$T$, $Q$, $U$ maps at 95\,GHz
using data taken by \bicepthree\ during the 2016--2018
seasons---we refer to these maps as BK18$_{B95}$.
The left column shows the real data maps with $0.25\deg$ pixelization
as output by the reduction pipeline.
The right column shows a noise realization made by randomly assigning
positive and negative signs while coadding the data.
These maps are filtered by the instrument beam
(24~arcmin FWHM~\cite{biceptwoXV}),
timestream processing, and (for $Q$~\&~$U$) deprojection of
beam systematics.
Note that the horizontal/vertical and $45\deg$ structures seen in the
$Q$ and $U$ signal maps are expected for an \emode\ dominated sky.}
\label{fig:tqu_maps_95}
\end{figure*}

\begin{figure*}
\resizebox{\textwidth}{!}{\includegraphics{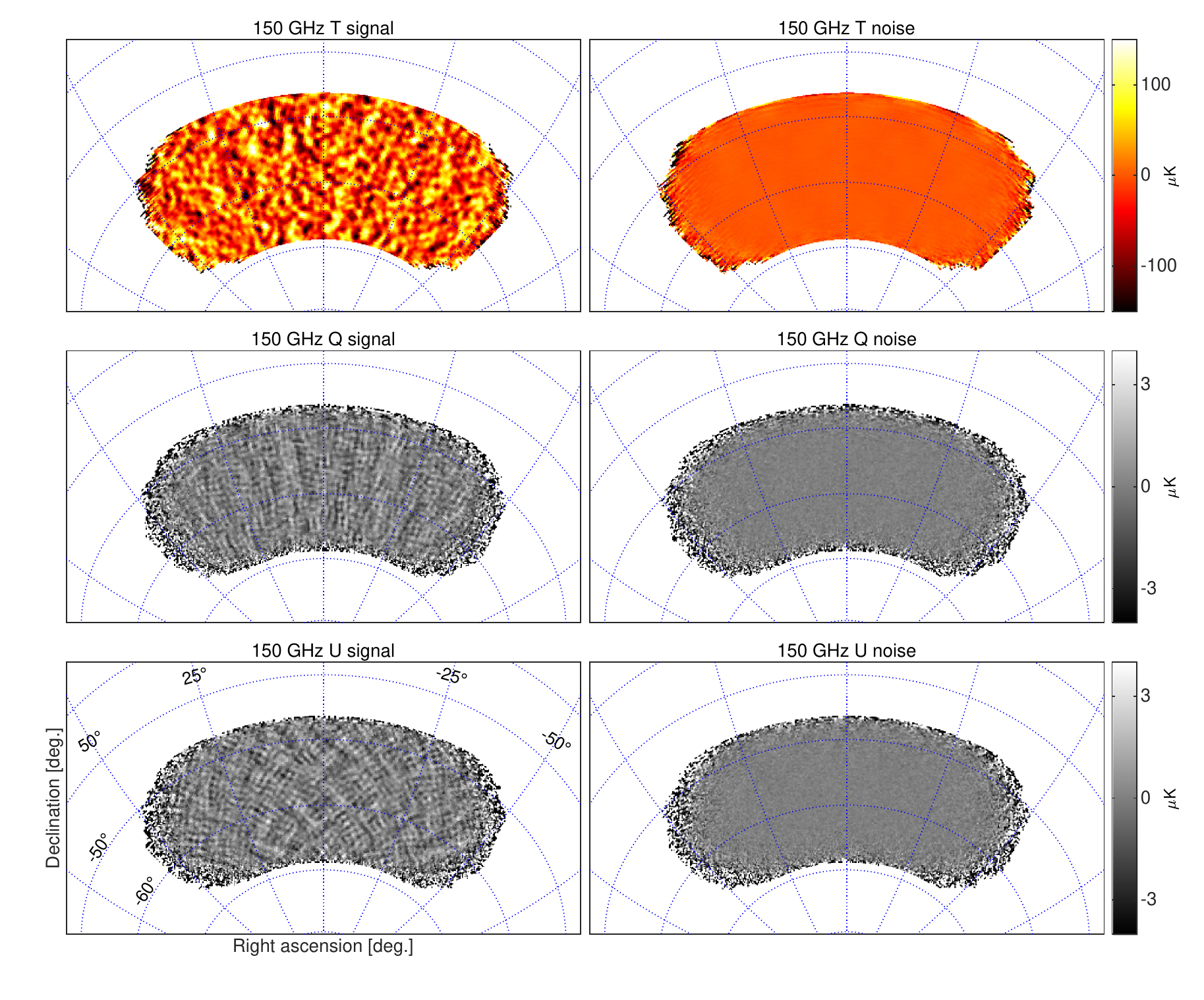}}
\caption{$T$, $Q$, $U$ maps at 150\,GHz
using data taken by \biceptwo\ and \keckarray\ during the 2010--2016
seasons---we refer to these maps as BK18$_{150}$.
These maps are directly analogous to the 95\,GHz maps shown in
Fig.~\ref{fig:tqu_maps_95} except that the
instrument beam filtering is in this case 30~arcmin FWHM~\cite{biceptwoXI}.
The smaller coverage region is due to the
smaller instantaneous field of view of the \biceptwo/\keck\ receivers.}
\label{fig:tqu_maps_150}
\end{figure*}

\begin{figure*}
\resizebox{\textwidth}{!}{\includegraphics{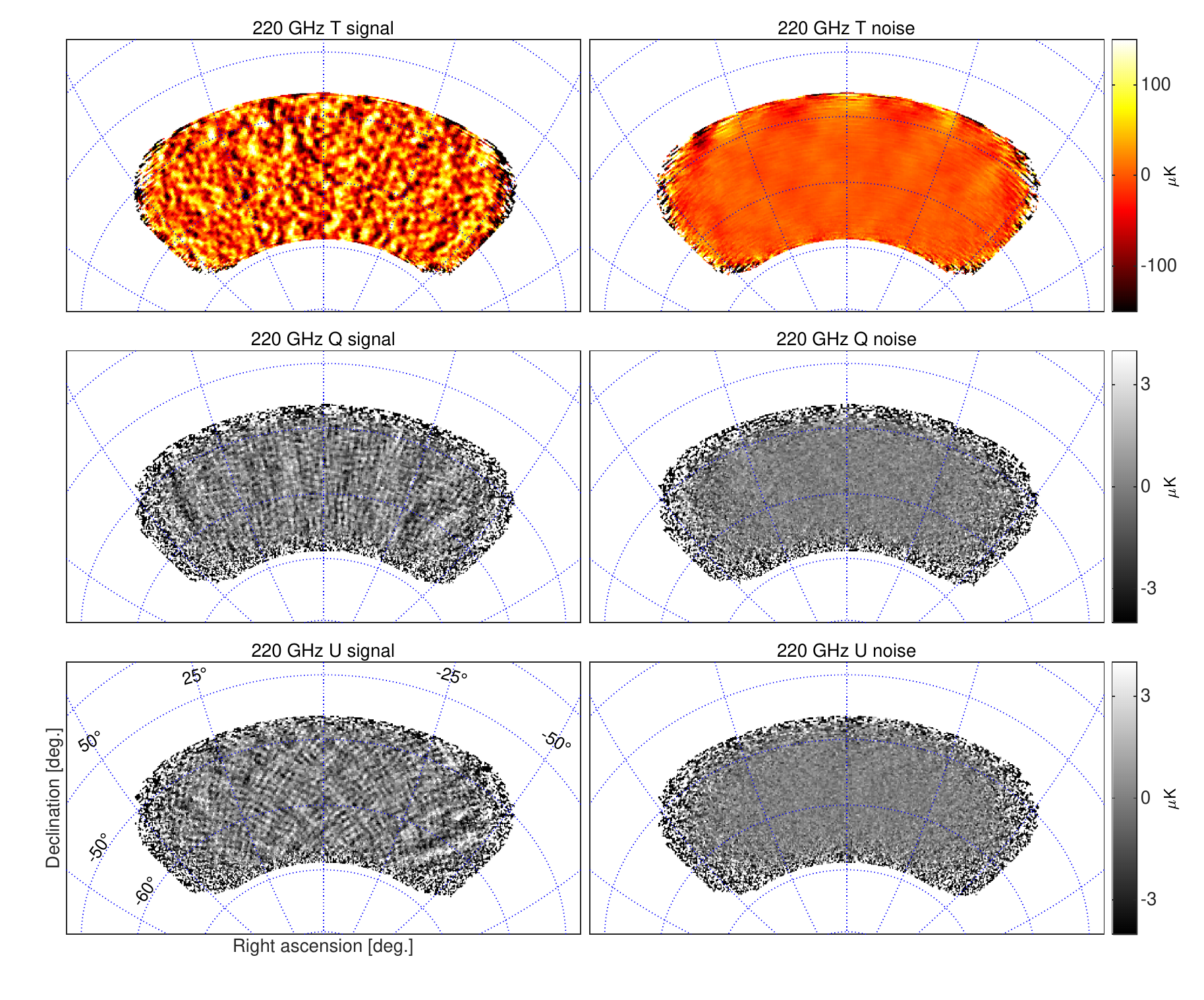}}
\caption{$T$, $Q$, $U$ maps at $\approx220$\,GHz
using 14 receiver-years of data taken by \keckarray\ during the 2015--2018
seasons---we refer to these maps as BK18$_{220}$.
These maps are directly analogous to the 95\,GHz maps shown in
Fig.~\ref{fig:tqu_maps_95} except that the
instrument beam filtering is in this case 20~arcmin FWHM~\cite{biceptwoXI}.
Note the polarized dust emission which is visible in
the right part of the $Q/U$ signal maps.}
\label{fig:tqu_maps_220}
\end{figure*}

Figures~\ref{fig:tqu_maps_95},~\ref{fig:tqu_maps_150}~\&~\ref{fig:tqu_maps_220}
show $T$/$Q$/$U$ maps at 95\,GHz from \bicepthree, at
150\,GHz from \biceptwo/\keck\ and at 220\,GHz from \keck.
The right side of each figure shows realizations of noise
created by randomly flipping the sign of data subsets while
coadding the map---see Sec.~V.B of Ref.~\cite{biceptwoI} for
further details.
Due to the larger instantaneous field of view \bicepthree\ naturally
maps a larger area of sky.
We have perturbed the boresight ``scan-box'' on the sky to
higher declination such that most of the additional area lies
on that side of the \biceptwo/\keck\ observed region.
Comparing the three frequencies we see almost
identical patterns in the $T$ maps since \lcdm\ signal
is dominant.
For $Q$ and $U$ the patterns are again extremely similar
between 95 and 150\,GHz, and dominated by the \lcdm\ \emode\
signal.
However, at 220\,GHz we start to see dust contamination become
comparable in amplitude to the \lcdm\ {\emode}s at the edges
of the field.

\section{Internal Consistency Tests of New Data}
\label{app:mapjack}

A powerful internal consistency test are map level data split difference tests
which we refer to as ``jackknives''.
As well as the full coadd signal maps we also form many pairs
of split maps where the splits are chosen such that one might
expect different systematic contamination in the two halves
of the split.
The split halves are differenced and the power spectra
taken.
We then take the normalized deviations of the real bandpowers from the mean of
signal+noise simulations and
form $\chi^2$ and $\chi$ (sum of deviations) statistics.
In this section we perform tests of the new 2016, 2017
and 2018 data which are the same as those performed on the
previous \bicep/\keck\ data in
Sec.~VII.C of Ref.~\cite{biceptwoI}, Sec.~6.3 of Ref.~\cite{biceptwoV},
Appendix~B of BK14, and Appendix~B of BK15.
In this paper we are adding three new years of data and
we choose to test each band/year separately.
The fourteen data splits themselves remain as described previously
and we again do each $\chi^2$ and $\chi$ test twice using
the lowest five and lowest nine bandpowers ($\ell<200$ and $\ell<300$),
and for each of the $EE$, $BB$ and $EB$ spectra.

\begin{figure}[htb]
\begin{center}
\resizebox{0.7\columnwidth}{!}{\includegraphics{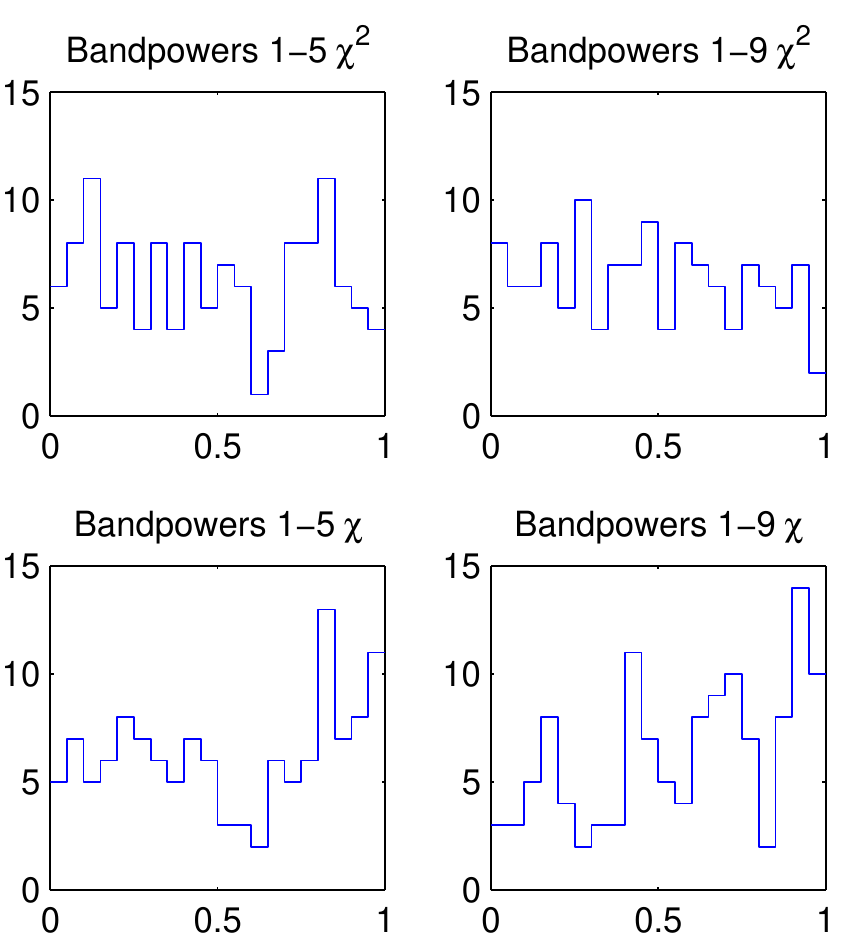}}
\end{center}
\caption{Distributions of the jackknife $\chi^2$ and $\chi$ PTE
values for the \bicepthree\ 2016, 2017, \& 2018 95\,GHz data.
This figure is analogous to Fig.~10 of BK15.}
\label{fig:ptedist_95}
\end{figure}

\begin{figure}[htb]
\begin{center}
\resizebox{0.7\columnwidth}{!}{\includegraphics{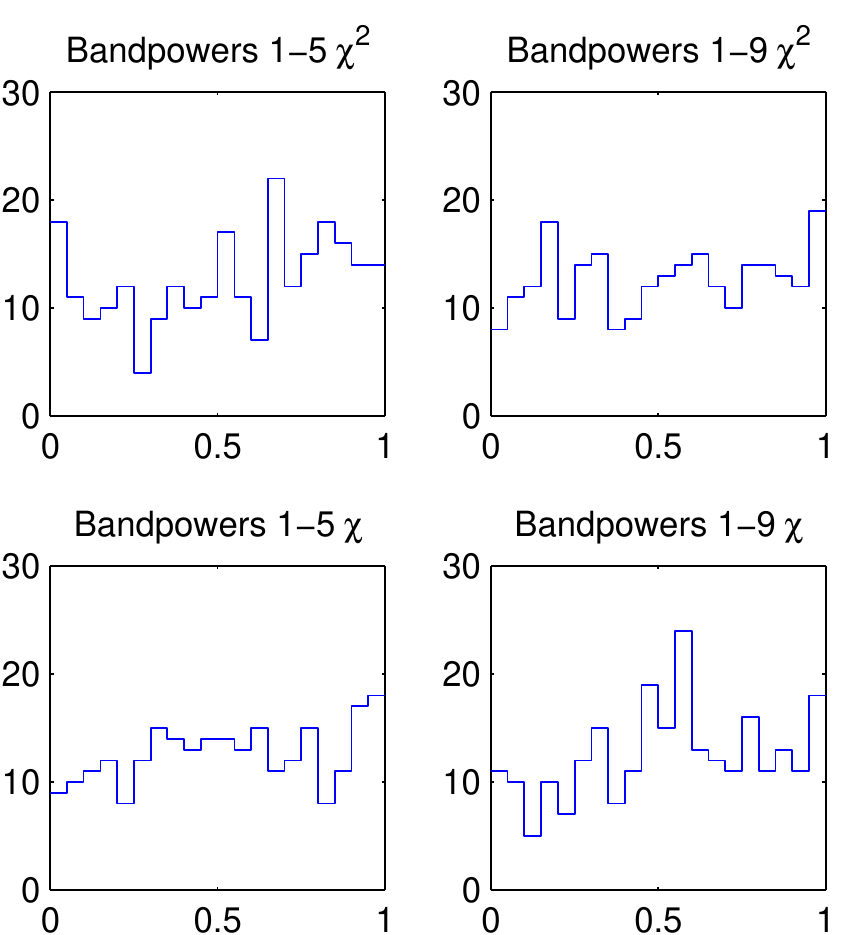}}
\end{center}
\caption{Distributions of the jackknife $\chi^2$ and $\chi$ PTE
values for the \keckarray\ 2016, 2017, \& 2018 220\,GHz data.
This figure is analogous to Fig.~11 of BK15.}
\label{fig:ptedist_220}
\end{figure}

The two nominally 220\,GHz receivers added before the 2015 season
actually have passbands centered at $\approx 230$\,GHz,
whereas the two added before the 2016 season have passbands
centered at $\approx 220$\,GHz.
We made separate maps using these receivers and perform
separate jackknife tests on these.
(We have chosen to coadd these sub-bands in the main
analysis to reduce the number of cross-spectra.)
For 95\,GHz we have $3$ years $\times 14$ data-splits $\times 2$
statistics ($\chi^2$ or $\chi$) $\times 2$ bandpower-ranges
$\times 3$ spectra ($EE$/$BB$/$EB$) $= 504$ tests, and for 220/230\,GHz
twice as many again for 1008 tests.
The $\chi^2$ and $\chi$ jackknife statistics do not actually
have the nominal theory distributions since they are summations of bandpower values
which are themselves $\chi^2$ rather than Gaussian distributed.
We deal with this by computing PTE values versus the simulations
which should have the correct distributions.
Since there are now more tests than simulation realizations (499)
we expect a few zero and unity PTE values---there
are 2 for \bicepthree\ and 3 for \keck\ 220/230\,GHz.
However, we note that in these cases the real data value is only
just beyond the end of the simulated distribution, and when
compared to the distribution of simulated values aggregated
over all jackknifes the values are within the expected range.
Figures~\ref{fig:ptedist_95}~\&~\ref{fig:ptedist_220} show
histograms of the PTE values breaking down into the $\chi^2$ and $\chi$
tests over the two bandpower ranges.
Problems with the noise debias and/or estimated fluctuation
would be expected to result in an excess number of values
in the edge bins.

In the initial pass of the jackknife analysis there were
a few $\chi^2$ and $\chi$ values sitting sufficiently beyond the tails of the simulation
distributions that it was unlikely to happen by chance.
Through a long process we determined that removing two small
sub-sets of data moved these values in such that there were no
longer clear problems.
Mostly data is not taken during the ``station open'' period from
early November through mid-February, and when it has been it has typically
not been included in science analysis.
In 2016 data was taken with \keckarray\ for the full month
of November and initially this was included in the analysis.
However, we found evidence for failure of the season split jackknife
when including this data which lead us to exclude it.
In addition we found it necessary to remove \bicepthree\ data
from focal plane tile~1 in order to pass the tile jackknife.
Due to readout limitations the partner tile across the boresight
was not populated and was instead blanked off with a reflective
plate.
The hypothesis is that this led to reflections and ghost beams
for these detectors---see Ref.~\cite{biceptwoXV} for further
details.
We note that we did not look at the non-jackknife spectra until
the above data removals had been done and the
null tests determined to be satisfactory.

\section{95 \& 220\,GHz Spectral Stability}
\label{app:specjack}

\begin{figure}[htb]
\begin{center}
\resizebox{\columnwidth}{!}{\includegraphics{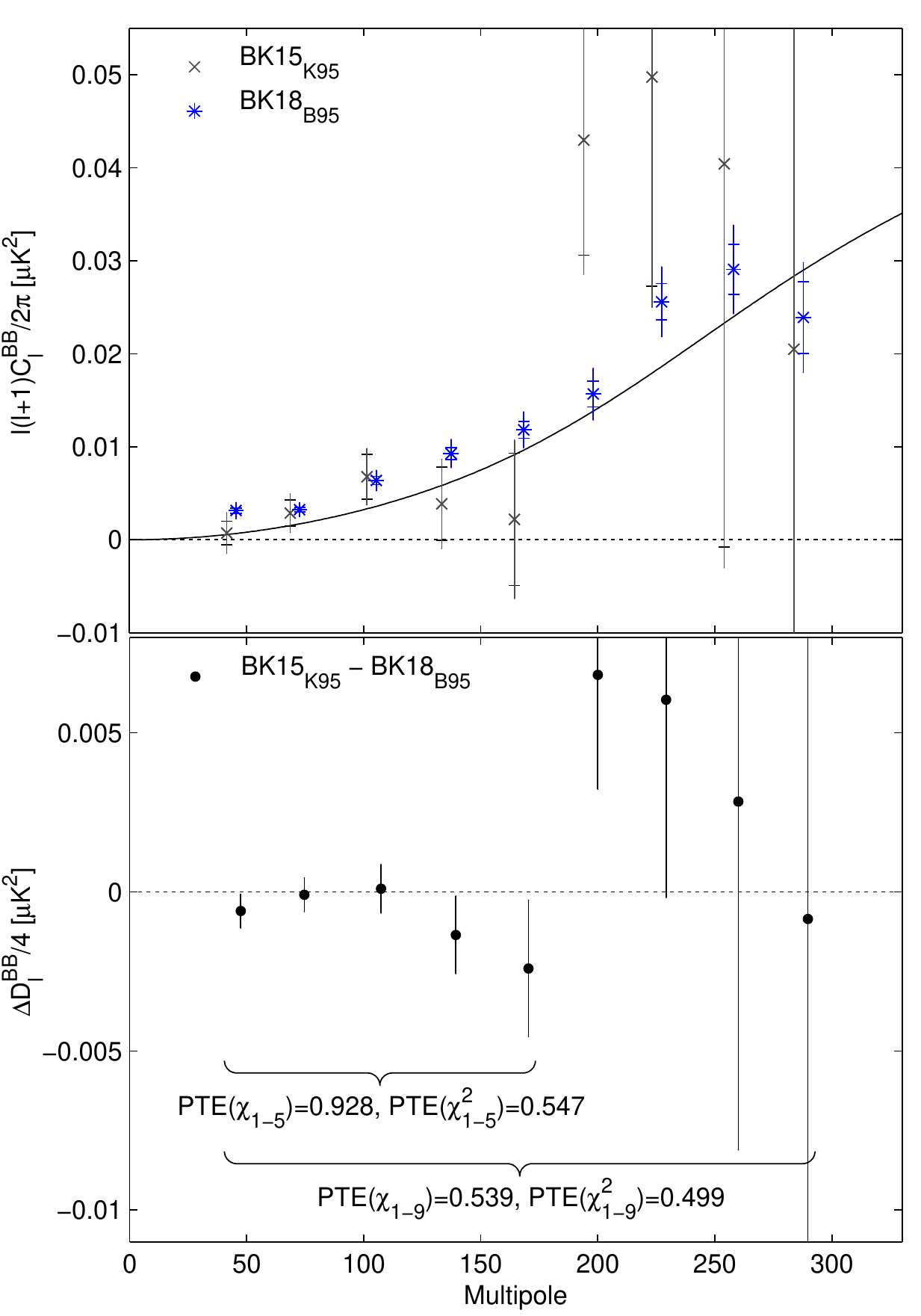}}
\end{center}
\caption{{\it Upper:} Comparison of the previously published
95\,GHz $BB$ auto-spectrum from the \keckarray\ receivers (BK15$_{\mathrm{K95}}$) to
the new \bicepthree\ spectrum (BK18$_{\mathrm{B95}}$).
The \bicepthree\ sky coverage region is a super-set of the
\keck\ region.
The inner error bars are the standard deviation of the noise
realizations, while the outer error bars also include signal sample
variance (lensed-\lcdm+dust).
Neither of these uncertainties are appropriate for comparison
of the band power values---for this see the lower panel.
The solid line shows the lensing \bmode\ expectation.
(For clarity the sets of points are offset horizontally.)
{\it Lower:} The difference of the spectra shown
in the upper panel divided by a factor of four.
The error bars are the standard deviation of the pairwise
differences of signal+noise simulations
which share common input skies (the simulations used to
derive the outer error bars in the upper panel).
Comparison of these points with null is an appropriate test
of the compatibility of the spectra, and the PTE of $\chi$
and $\chi^2$ are shown.
This figure is similar to Fig.~12 of BK15.}
\label{fig:specjack_95}
\end{figure}

We next test the mutual compatibility of the previously
published BK15 spectra and the new BK18 spectra.
In 2014 and 2015 \keckarray\ operated with two receivers
at 95\,GHz producing a relatively low sensitivity map.
In 2016, 2017 and 2018 \bicepthree\ observed in the same
frequency band over an overlapping but somewhat larger
sky area.
The upper part of Fig.~\ref{fig:specjack_95} compares
the two $BB$ spectra---the large improvement in sensitivity
is clear.
Since the sky regions overlap the bandpowers are correlated.
To gauge consistency we compare the differences of the
real bandpowers to the differences of simulations which share
common input skies.
The lower part of the figure shows that this (not very stringent)
test is passed.

\begin{figure}[htb]
\begin{center}
\resizebox{\columnwidth}{!}{\includegraphics{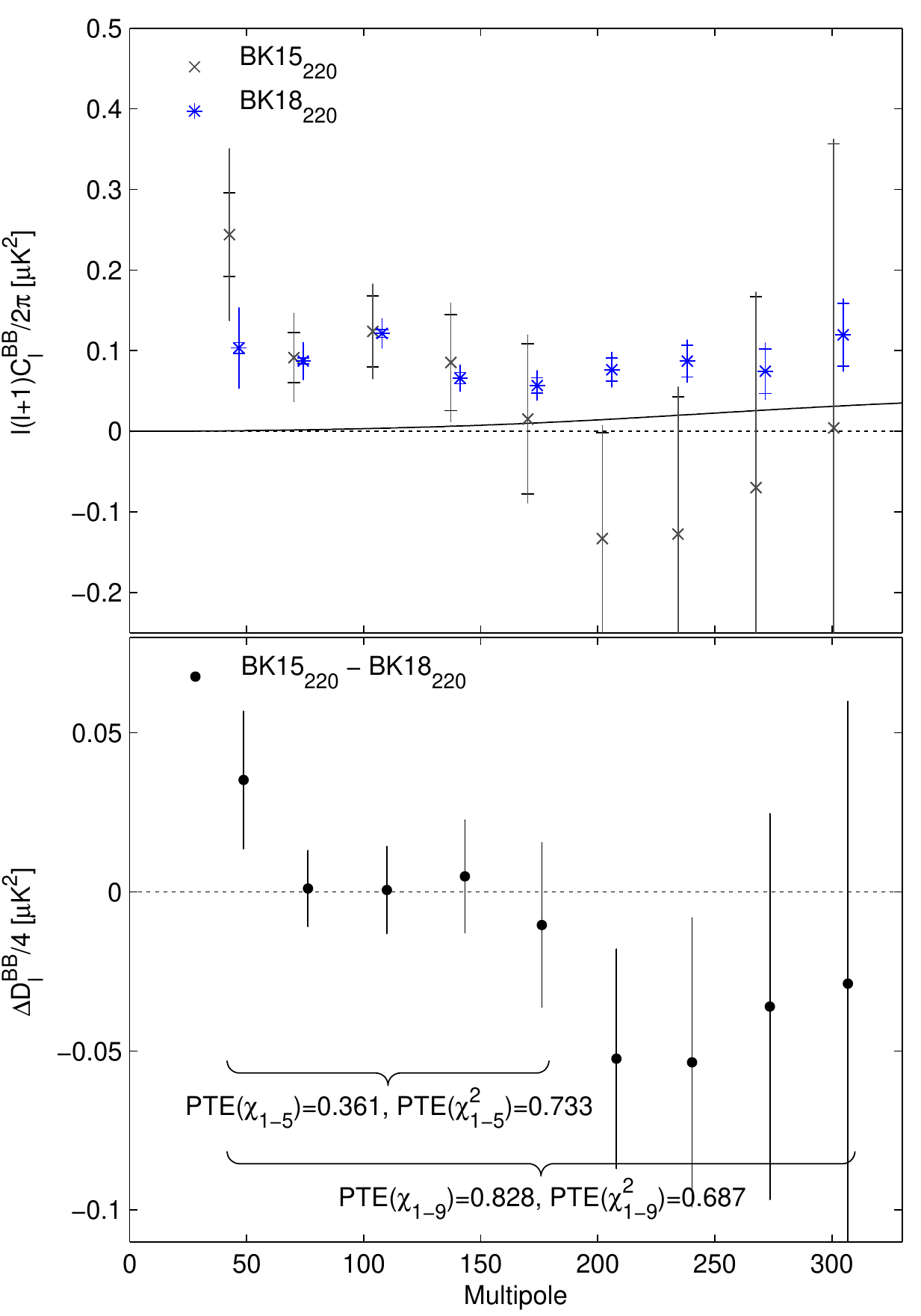}}
\end{center}
\caption{Comparison of the previously published 220\,GHz $BB$ auto-spectrum
(BK15$_{220}$) to the new BK18 version (BK18$_{220}$).
This figure is analogous to Fig.~\ref{fig:specjack_95}.
In this case the sky coverage regions fully overlap
but there is a small amount of common data between the two maps
(2 receiver-years of the 14 receiver-years in BK18).}
\label{fig:specjack_220}
\end{figure}

The BK15 220\,GHz map came from two \keck\ receivers running in
2015.
Before the 2016 season we added two more receivers and ran
all four through 2016, 2017 and 2018.
The upper part of Fig.~\ref{fig:specjack_220} compares
the two $BB$ spectra---a large improvement in sensitivity
is again clear, and
the BK18 220\,GHz spectrum is sample variance limited out to $\ell\sim200$.
The lower panel again shows no evidence for inconsistency
between the two spectra.

\section{Multi Frequency Spectra}
\label{app:allspec}

Since the BKP paper our analyses have relied on comparing
a multi-component parametric model of lensed-\lcdm+foregrounds+noise
to the ensemble of auto- and cross-spectra between the \bicep/\keck\
and \wmap/\planck\ bands.
Through BK15 the coverage pattern of the \bicep/\keck\ bands on
the sky was very similar.
The apodization mask chosen was the geometric mean of the
inverse noise variance maps, and this was also applied to the \wmap/\planck\
bands.
\bicepthree\ maps a larger area of sky as can be seen comparing
Figures~\ref{fig:tqu_maps_95} and~\ref{fig:tqu_maps_150}.
In this paper we define the baseline BK18 analysis as being one where
the \bicepthree\ maps are apodized by their ``natural'' larger
inverse noise variance pattern, and
the \wmap/\planck\ maps are also apodized with this pattern.
Using different apodizations suppresses the amplitude of the cross-spectra
as compared to the auto-spectra but this is automatically taken
into account in the bandpower window function calculations~\cite{biceptwoVII,willmert20}.
Using different apodizations also in principle reduces the correlations
of the auto- and cross-spectra which are key to the separation of
foreground and CMB signals.
However, in the current case the penalty turns out to be negligible.
We consider some alternate apodizations in Appendix~\ref{app:likevar} below.

\begin{figure*}
\begin{center}
\resizebox{\textwidth}{!}{\includegraphics{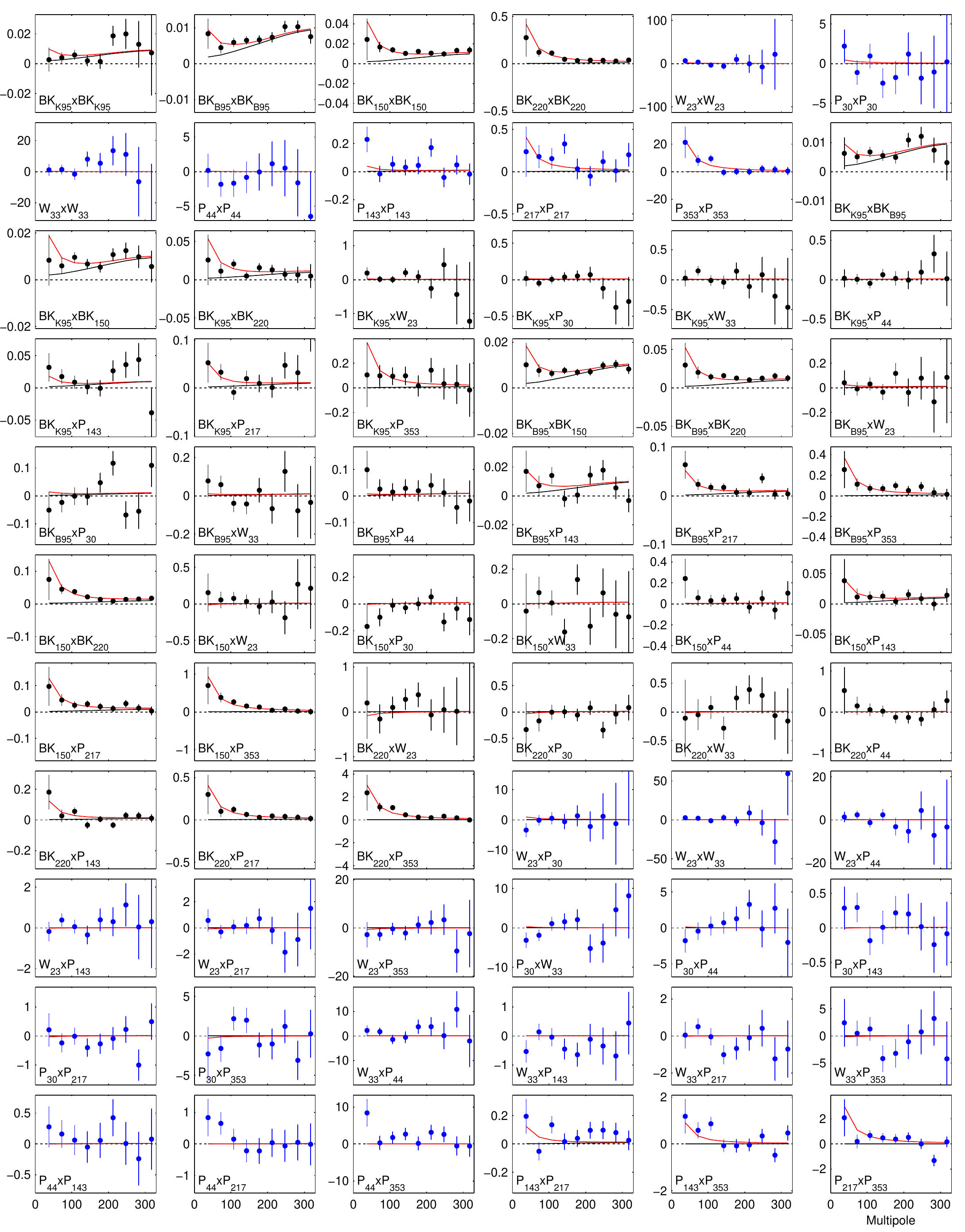}}
\end{center}
\caption{The full set of $BB$ auto- and cross-spectra
from which the joint model likelihood is derived.
In all cases the quantity plotted is $100 \ell C_\ell / 2 \pi$ (\uksq).
Spectra involving \bicep/\keck\ data are shown as black points
while those using only \wmap/\planck\ data are shown as blue
points.
The black lines show the expectation values
for lensed-\lcdm, while the red lines show the expectation values of the
maximum likelihood lensed-\lcdm+$r$+dust+synchrotron model
($r=\rmlm$, $\Adf=\Admlm$\,\uksq, $\Bd=\Bdmlm$, $\ad=\admlm$,
 $\Asf=\Asmlm$\,\uksq, $\Bs=\Bsmlm$, $\as=\asmlm$, $\epsilon=\emlm$),
and the error bars are scaled to that model.}
\label{fig:powspec_all}
\end{figure*}

Fig.~\ref{fig:powspecres_bkbands}
shows only a small subset of the spectra which are used in the
likelihood analysis and included in the provided \texttt{COSMOMC}
file.
In this paper we use four \bicep/\keck\ maps, two \wmap\ maps (23 and 33\,GHz),
and five \planck\ maps (30, 44, 143, 217 and 353\,GHz) resulting
in 11 auto- and 55 cross-spectra.
(To keep down the total number of spectra we drop the other \planck\
bands since these provide a negligible amount of additional information.)
In Fig.~\ref{fig:powspec_all} we show all of the included spectra
together with the maximum likelihood model from the baseline analysis
whose parameters were quoted above.
Most of the spectra not already shown in Fig.~\ref{fig:powspecres_bkbands}
have low signal-to-noise, although a few of them carry interesting
additional information on the possible level of synchrotron.

To quantify the absolute goodness-of-fit of the data
to the maximum likelihood model we form statistics from
the real data bandpowers and compare these to corresponding
values from the simulation realizations.
For the $9\times66=594$ bandpowers shown in Fig.~\ref{fig:powspec_all},
$\chi^2=(d-m)^T C^{-1}(d-m)=\chitwo$, where $d$ are the
bandpower values, $m$ are the model expectation values,
and $C$ is the bandpower covariance matrix for the maximum likelihood model.
This has a PTE versus the simulations of $\chitwoptesim$.
If instead we take the sum of the normalized deviations
$\chi = \sum{((d-m)/e)}$ where $e$ is the square-root of the
diagonal of $C$, we find that the PTE versus the simulations
is \chioneptesim.
The parametric model which we are using, including
the approximation of Gaussian fluctuation
of the dust (and synchrotron) sky patterns, remains an adequate
description of the presently available data.

\begin{figure}
\begin{center}
\resizebox{\columnwidth}{!}{\includegraphics{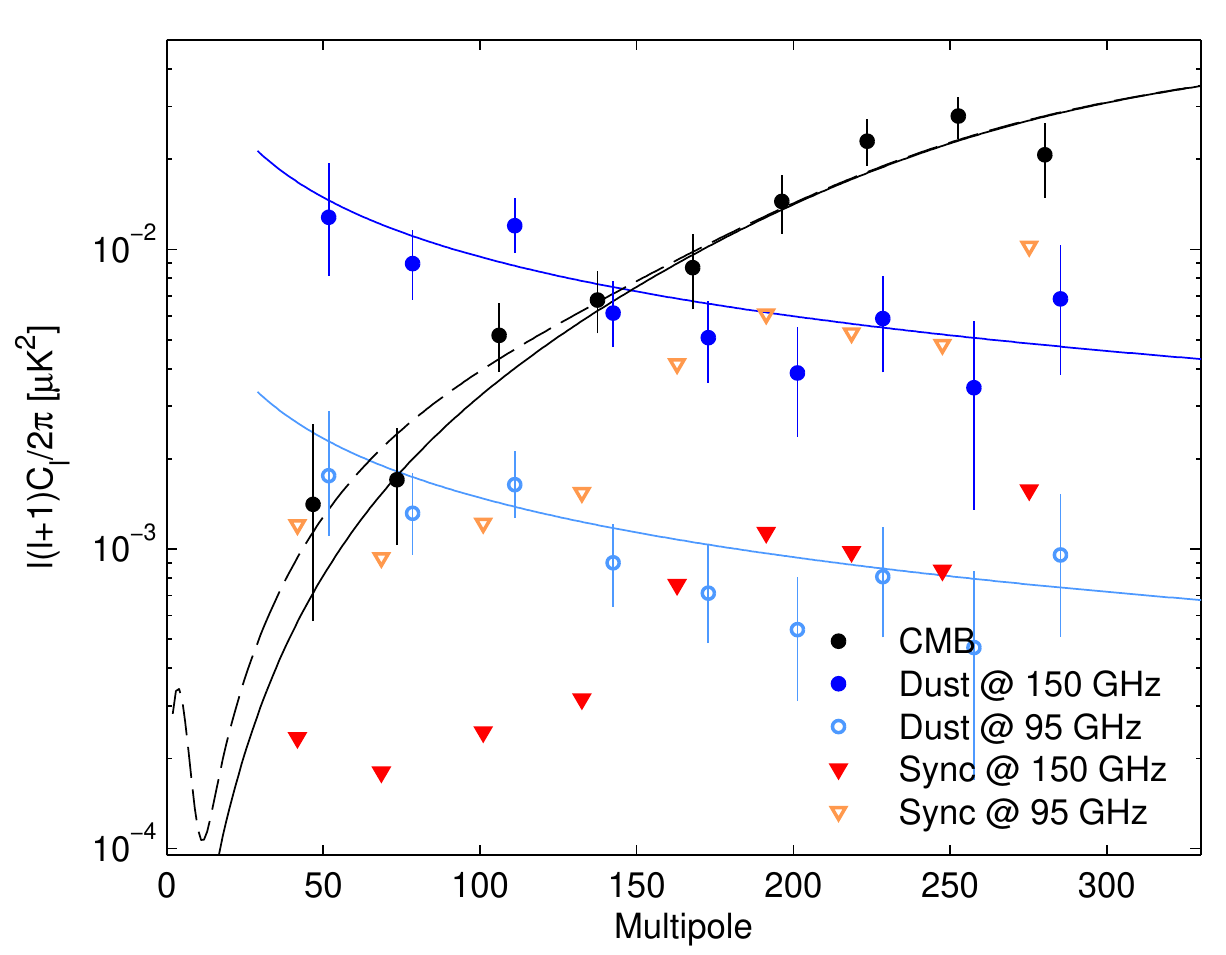}}
\end{center}
\caption{Spectral decomposition of the $BB$ data into synchrotron
(red), CMB (black) and dust (blue) components at 150\,GHz (filled points)
or 95\,GHz (open points).
The decomposition is calculated independently in each bandpower,
marginalizing over $\Bd$, $\Bs$ and $\epsilon$, with priors.
Error bars denote 68\protect\% credible intervals, with the point marking the most
probable value.
For synchrotron, which is not detected in this analysis, the downward triangles
denote 95\% upper limits.
(For clarity the sets of points are offset horizontally.)
The solid black line shows lensed-\lcdm\ with the dashed
line adding on top an $r_{0.05}=\rmlm$ tensor contribution.
The blue curves show the maximum-likelihood dust model from the baseline analysis
($\Adf=\Adcentval$\,\uksq, $\Bd=\Bdmlm$, and $\ad=\admlm$) as evaluated
at 150\,GHz (dark blue) or 95\,GHz (light blue).
}
\label{fig:specdecomp}
\end{figure}

As in BK15 we also run a likelihood analysis
to find the CMB and foreground contributions on
a bandpower-by-bandpower basis.
The baseline analysis is a single fit to all 9 bandpowers
across 66 spectra with 8 parameters.
Instead we now perform 9 separate
fits---one for each bandpower---across the 66 spectra,
with 6 parameters in each fit.
These 6 parameters are the amplitudes of CMB,
dust and synchrotron plus $\Bd$, $\Bs$ and $\epsilon$.
While the Planck-derived $\Bd$ prior ($\Bd=1.59\pm0.11$) is no longer
necessary for the baseline analysis, it is needed for this bin-by-bin analysis
which otherwise does not have enough information to constrain the dust
spectrum in the higher $\ell$ bins.
We repeat the analysis with foreground amplitudes defined at 150\,GHz, where
most of the \biceptwo\ and \keckarray\ sensitivity is concentrated, and
at 95\,GHz, the \bicepthree\ observing frequency.
The results are shown in Fig.~\ref{fig:specdecomp}---the
resulting CMB values are consistent with lensed-\lcdm\
while the dust values are consistent with the level
of dust found in the baseline analysis.
Synchrotron is tightly limited in all the multipole ranges,
and not detected in any of them.
We note that this figure offers empirical backing for an
assumption which we make in our mainline analysis:
that the power spectrum of the polarized dust emission
in our field follows a power law.

\section{Likelihood Variation and Validation}
\label{app:likeevolvar}

\subsection{Likelihood Evolution}
\label{app:likeevol}

\begin{figure*}[htb]
\begin{center}
\resizebox{1.0\textwidth}{!}{\includegraphics{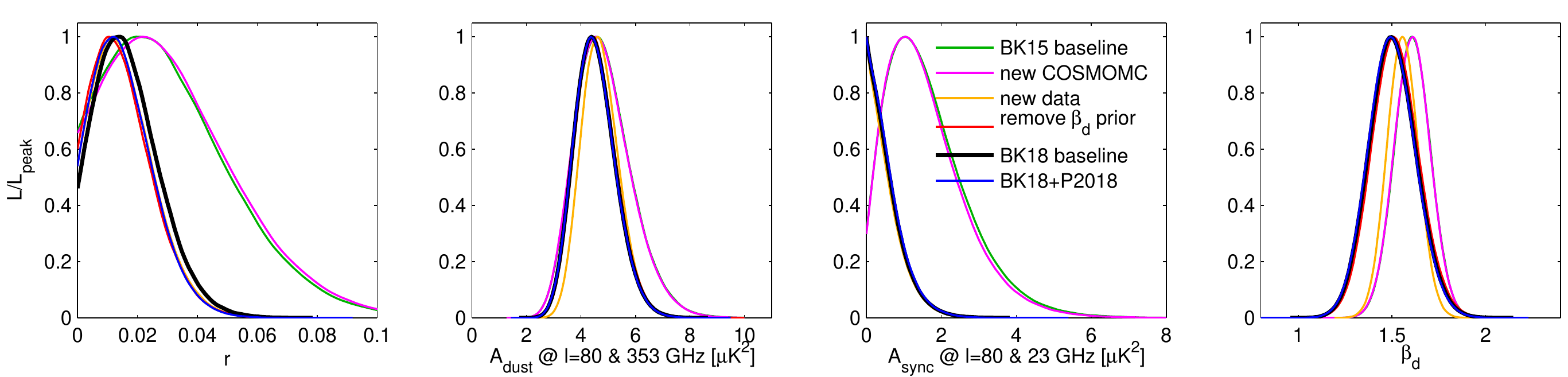}}
\end{center}
\caption{Evolution of the BK15 analysis to the new BK18 ``baseline'' analysis as
defined in this paper.
See Appendix~\ref{app:likeevol} for further details.}
\label{fig:likeevol}
\end{figure*}

Fig.~\ref{fig:likeevol} shows the sequence of steps from
the BK15 baseline analysis to the new baseline.
Starting with the BK15 data set we first update the version of
COSMOMC from Nov2016 to July2019 (green to magenta) which makes very little
difference.
We next switch from the BK15 data set to the BK18 one---i.e.
including the improved sensitivity at 220\,GHz and the improved
sensitivity and sky area at 95\,GHz (magenta to yellow).
This makes a substantial difference: the $r$ constraint
narrows and its peak position shifts down.
In addition the synchrotron amplitude constraint $\As$ shifts
down to peak at zero.

The BKP, BK14 and BK15 analyses all placed a prior on the
frequency spectral index of dust of $\Bd=1.59 \pm 0.11$, using a Gaussian prior
with the given $1\sigma$ width, this being an upper limit
on the patch-to-patch variation~\cite{planckiXXII,bkp}.
In BK15 the need for this prior was becoming weaker with the
data able to constrain $\Bd$ to $1.65 \pm 0.20$.
With the improved sensitivity of BK18 the constraint becomes
$\Bd=\Bdrange$.
In Fig.~\ref{fig:likeevol} dropping the $\Bd$ prior corresponds to
the transition from yellow to red curves---$r$ and $\As$ show very little change
while $\Ad$ gets just slightly wider.
Since there is little downside we choose to drop the $\Bd$ prior in the baseline
analysis for BK18, gaining a little confidence that we are not inducing any
bias due to the value of $\Bd$ in our particular sky patch differing from
the central value of the prior.
Interestingly the peak value of $\Bd=1.50$ selected by our data is
close to the mean value from \planck\ 2018 analysis of larger regions of sky
${\Bd^P=1.53\pm0.02}$~\cite{planck2018XI}.

The BKP, BK14 and BK15 likelihood analyses all assumed a lensing \bmode\
spectrum based on Planck 2013 cosmological parameters.
The predicted lensing $BB$ spectrum for the Planck 2018
baseline cosmological parameters~\cite[Table 2]{planck2018VI} has shifted down by
5\% (for $\ell<500$).
In Fig.~\ref{fig:likeevol} switching to the updated Planck 2018 spectrum
corresponds to the transition from red to black---the $r$ constraint
shifts slightly up indicating that this is a marginally relevant change
at BK18 sensitivity levels.
We define this analysis as the ``BK18 baseline'' and it
is this that is shown in Fig.~\ref{fig:likebase}.

Since there is residual uncertainty in the \planck\ 2018
cosmological parameters there is also residual uncertainty in the
prediction of the \lcdm\ lensing $BB$ spectrum.
In a simple study we find this to be 2.2\% within the multipole
range $20<\ell<500$ (with a distribution close to Gaussian).
To marginalize over this uncertainty we run a joint
analysis of the \planck\ 2018 TT,TE,EE+lowE+lensing likelihood and
the BK18 likelihood producing the shift in Fig.~\ref{fig:likeevol}
from heavy black to blue.
By doing this we also include the small additional constraining
power on $r$ which comes from the \planck\ TT,TE,EE+lowE+lensing likelihood.
It is this joint likelihood which is shown in Fig.~\ref{fig:rns}.

\subsection{Likelihood Variation}
\label{app:likevar}

\begin{figure*}[htb]
\begin{center}
\resizebox{1.0\textwidth}{!}{\includegraphics{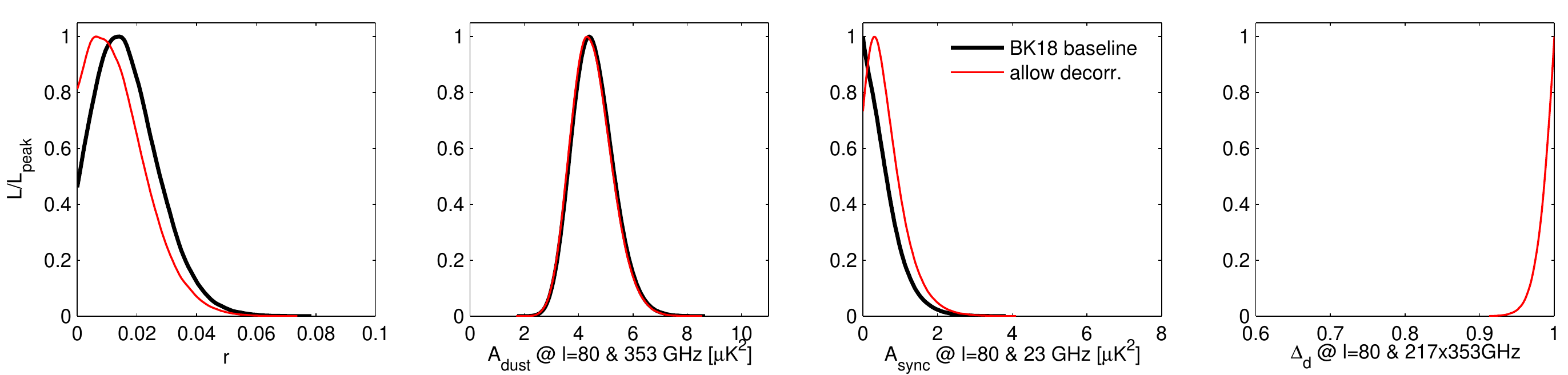}}
\end{center}
\caption{
Likelihood results when freeing dust decorrelation parameter $\dd$.
See Appendix~\ref{app:likevar} for details.}
\label{fig:likedecorr}
\end{figure*}

{\it Dust Decorrelation.}---Spatial or line-of-sight variations
of $\Bd$ will produce variations in the dust
polarization pattern with frequency.
This will result in suppression of the cross-spectra with respect
to the geometric mean of the auto-spectra---a phenomenon referred to
as decorrelation.
The \planck\ 2018 dust analysis~\citep{planck2018XI} did not show
any evidence for dust decorrelation.
Recently Ref.~\cite{pelgrims21} reported evidence for line-of-sight
decorrelation in \planck\ data by using HI data to select
sight lines which integrate over multiple dust clouds.
This study uses maps of the effective number of clouds taken
from Ref.~\cite{panopoulou20} which in Sec.~4.2 specifically examines
the \bicep/\keck\ patch and states that ``one cloud dominates the column
density in the majority of pixels in the \bicep/\keck\ region''.
Nonetheless, decorrelation surely exists in our sky patch at some level---the
question is whether it is relevant as compared to the
BK18 experimental noise.
In Appendix~F of BK15 we added a decorrelation parameter $\dd$ to
our parametric model, defined as the ratio of the cross-spectrum
$217\times353$ to the geometric mean of the corresponding auto-spectra,
and varying with frequency in a manner suggested by Ref.~\cite{planckiL}.
Fig.~\ref{fig:likedecorr} shows the result of allowing
this parameter to vary freely (for the case where $\dd$ is assumed to scale as $\ell/80$).
We see that the $r$ constraint shifts down slightly and the $\As$
constraint shifts up slightly.
In contrast to Fig.~21 of BK15 the $\dd$ curve peaks at unity
(no decorrelation).
Note that $\dd$ is a biasing parameter with respect to $r$---in BK15
it was found that allowing $\dd$ to vary freely resulted in 72\%
of the $r$ curves peaking at zero in simulations where there was no
decorrelation.
We therefore do not allow $\dd$ to vary freely in our baseline analysis at
this time.
(To avoid having to define dust decorrelation and dust/sync
correlation simultaneously when we allow $\dd$ to vary freely we fix
$\epsilon=0$.)

\begin{figure*}[htb]
\begin{center}
\resizebox{1.0\textwidth}{!}{\includegraphics{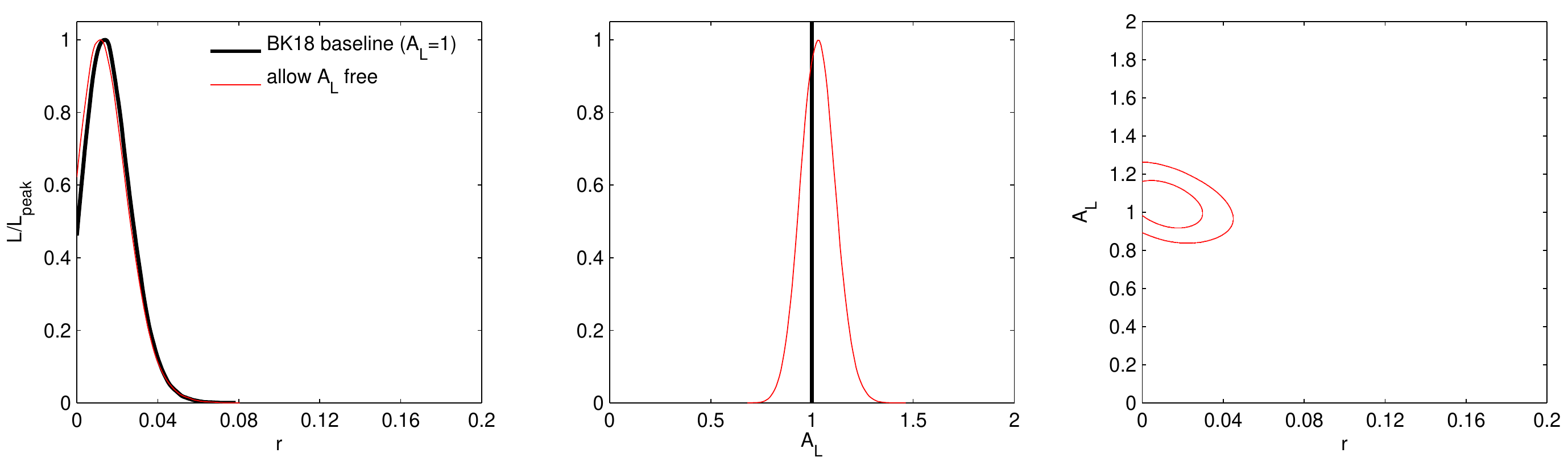}}
\end{center}
\caption{
Likelihood results when allowing the lensing amplitude
to be a free parameter---see Appendix~\ref{app:likevar} for details.}
\label{fig:freelens}
\end{figure*}

{\it Free Lensing Amplitude.}---The \planck\ 2018 cosmological parameters
predict a specific lensing $BB$
spectrum as used in the ``BK18 baseline'' analysis described above.
If we introduce an artificial scaling parameter on this spectrum
$A_{\rm L}^{\rm BB}$ and repeat the analysis we get
the results shown in Fig.~\ref{fig:freelens}.
With a uniform prior on $A_{\rm L}^{\rm BB}$
(and marginalizing over all other parameters)
we obtain $A_{\rm L}^{\rm BB}= 1.03^{+0.08}_{-0.09}$ showing no tension with \lcdm.
The $r$ constraint hardly changes due to the fact that the lensing
amplitude prefered by the BK18 data is very close to that predicted
by the \planck\ 2018 cosmological parameters.
This is an interesting result given the \planck\ 2018 TT,TE,EE+lowE
result $A_{\rm L}=1.18\pm0.065$~\cite[Eqn.~36b]{planck2018VI}.
In addition in the right panel of Fig.~\ref{fig:freelens} we see
that there is very little degeneracy between $r$ and $A_{\rm L}^{\rm BB}$
implying that all one now needs from \lcdm\ is the template shape
of the lensing \bmode\ spectrum, and not the amplitude.

\begin{figure*}[htb]
\begin{center}
\resizebox{1.0\textwidth}{!}{\includegraphics{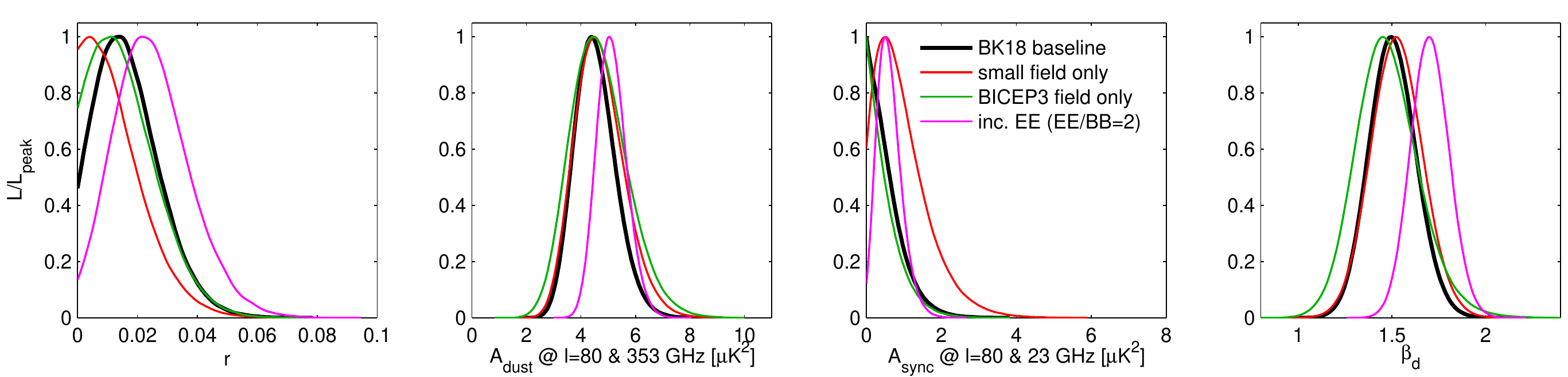}}
\end{center}
\caption{
Likelihood results when varying the internal data set selection.
The result including $EE$ should not be overinterpreted.
See Appendix~\ref{app:likevar} for details.
}
\label{fig:datavar}
\end{figure*}

{\it Sky Coverage.}---As seen in Figures~\ref{fig:tqu_maps_95} and~\ref{fig:tqu_maps_150}
the \bicepthree\ sky coverage is larger than the \biceptwo/\keck\ region.
As already mentioned in Appendix~\ref{app:allspec} for the baseline
BK18 analysis we apodize each map with its ``natural''
inverse variance coverage pattern and then take auto- and cross-spectra.
The \wmap\ and \planck\ maps are apodized with the \bicepthree\
coverage pattern.
Running maximum likelihood searches on simulations indicates that this
results in $\sigma(r)$ as low as any other option with no detectable
bias on $r$.
Fig.~\ref{fig:datavar} shows two alternatives:
``small field only'' apodizes all maps with the \biceptwo/\keck\ coverage
pattern and ``BICEP3 field only'' drops the \biceptwo/\keck\ 95, 150 and 220\,GHz maps
and just uses \bicepthree\ at 95\,GHz plus the \wmap/\planck\ bands.
Interestingly the peak position of $\Ad$ only shifts a very small amount under these
variations showing no evidence for variation
of the polarized dust amplitude as averaged across the \bicepthree\
coverage pattern versus the subset \biceptwo/\keck\ field.
It is very interesting that dropping the
150 and 220\,GHz bands does not substantially increase the width
of the $r$ constraint, although it is unsurprising that it does increase
the widths of the $\Ad$ and $\Bd$ constraints.
Note that for the small field only version the $\As$ peak shifts above
zero as it was in BK15.

{\it Including $EE$.}---Fig.~\ref{fig:datavar} also shows a case when including the $EE$ spectra
(and hence also the $EB$ spectra) under the artificial assumption
that the $EE/BB$ ratios for dust and synchrotron are both exactly 2,
as is shown to be close to the case in Refs~\cite{planck2018XI} and~\cite{krachmalnicoff18}.
The effects are similar to those seen in BK15---the $r$ peak position
shifts up, the $\Ad$ curve narrows, and the $\As$ curve peaks strongly away from zero.
It is unclear how much patch-to-patch variation
we should in fact allow in the $EE/BB$ ratio so these
results should not be overinterpreted at this time.

\begin{figure*}[htb]
\begin{center}
\resizebox{1.0\textwidth}{!}{\includegraphics{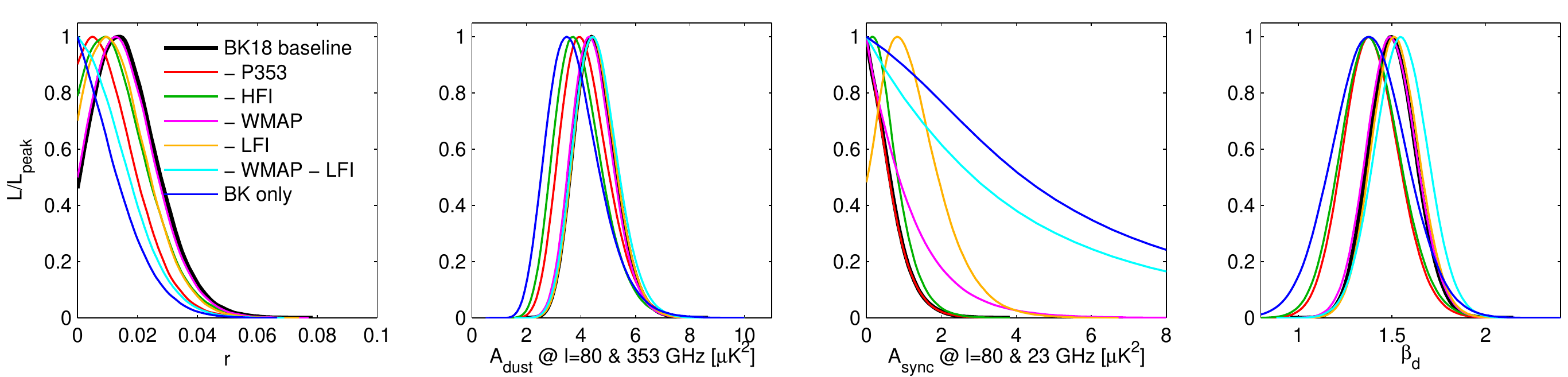}}
\end{center}
\caption{
Likelihood results when removing some or all of the external \wmap\ and
\planck\ bands---see
Appendix~\ref{app:likevar} for details.}
\label{fig:datavar_ext}
\end{figure*}

{\it External data variations.}---Fig.~\ref{fig:datavar_ext} shows variations
in the \wmap\ and \planck\ data being included.
With $\Bd$ now free, the \planck\ 353\,GHz band
provides a significant part of the $\Bd$ constraint (red curve
wider than black in the rightmost panel).
Dropping the other HFI bands makes smaller additional changes
(red to green).
It appears that there is some tension between the \wmap\ bands
(23 and 33\,GHz) and the LFI bands (30 and 44\,GHz) in terms
of their cross correlation with the \bicep/\keck\ maps.
Dropping \wmap\ results in $\As$ peaking strongly at zero whereas
dropping LFI results in $\As$ peaking well above zero.
Unsurprisingly dropping both \wmap\ and LFI results in a highly
degraded constraint on $\As$ (cyan).
\bicep/\keck\ alone (blue) constrains $\As$ and $\Bd$ poorly
with the $r$ constraint peaking hard at zero.

{\it Adding and removing fit parameters.}---In Appendix~\ref{app:allspec}
above we saw that the absolute
goodness-of-fit of the baseline model to the data is adequate.
In addition we can ask whether the data shows evidence
of the need for the baseline parameters, or preference for
additional ones, by considering $\Delta(\chi^2)$.
Performing maximum likelihood searches on the real data and taking
$-2\log(L)$ as a proxy for $\chi^2$, Table~\ref{tab:chi2} shows
the shifts which occur.
Allowing decorrelation parameter $\dd$ to vary freely while fixing
$\epsilon=0$ leaves the nominal number of parameters unchanged
and as expected hardly changes the mean value of $-2\log(L)$ for
the simulations.
The reduction of 0.3 for the real data offers no significant evidence
for detection of decorrelation.

\begingroup
\squeezetable
\begin{table}[pht]
\caption{\label{tab:chi2}
Values of $-2\log(L)$ (as a proxy for $\chi^2$) for the BK18 dataset
as the model is varied.
The last two rows are toy cases---see text for details.
}
\begin{ruledtabular}
\begin{tabular}{l r r r r r}
Model & mean sims & real data & $\Delta$ sims & $\Delta$ data & $\Delta$ nom. \\
\hline
baseline    &  589.65  &  535.99 &       &          &      \\
$+\dd-\epsilon$ &  589.73  &  535.69 &   0.08 &     -0.29 &     0 \\
$-$sync      &  592.19  &  537.52 &   2.54 &      1.52 &     4 \\
$-r$        &  590.71  &  536.58 &   1.06 &      0.59 &     1 \\
$-$sync$-r$    &  593.24  &  539.24 &   3.59 &      3.24 &     5 \\
$+A_L-\epsilon$ &  589.70  &  536.12 &   0.05 &      0.13 &     0 \\
$+A_L=0.5$  &  626.44  &  566.70 &  36.79 &     30.71 &     1 \\
$+\Ad=2$\,\uksq    &  602.55  &  553.80 &  12.90 &     17.81 &     1 \\
\end{tabular}
\end{ruledtabular}
\end{table}
\endgroup

Removing synchrotron from the fit (i.e.\ fixing $As=0$, $\Bs=0$,
$\as=0$ and $\epsilon=0$) corresponds to a nominal reduction in
the number of fit parameters of 4, but the increase
in the mean $-2\log(L)$ for the simulations (which contain no
synchrotron) suggests only 2.5 effective free parameters.
The increase of 1.5 for the real data is smaller than the average
increase in the simulations indicating no evidence that synchrotron
is present.

Fixing $r=0$ results in an increase of 1.0 for the mean of simulations
(which have $r=0$).
The increase for the real data is 0.6 indicating no significant evidence
for $r>0$.
Removing synchrotron and $r$ we are down to the 3 parameter model
($\Ad$, $\Bd$, $\ad$) from which the simulations were generated.
The mean of simulations responds exactly as the sum of the ``$-$sync'' and
``$-r$'' cases.
The real data shows an increase almost exactly equal to that in
the mean of simulations---i.e. the data is perfectly compatible with
the lensed-\lcdm\ expectation plus the simple 3 parameter dust model.

The $+A_L-\epsilon$ case shows no preference for varying the lensing scale
factor away from unity for either the real data (as expected given
the results in Fig.~\ref{fig:freelens} above) or the simulations.
The last two lines of the table are intended simply as test cases.
If we fix the lensing spectrum rescale factor $A_{\rm L}^{\rm BB}=0.5$
we see very large increases in $-2\log(L)$ for simulations and
data indicating strong incompatibility with such a model.
Likewise setting $\Adf=2$\,\uksq\ is strongly incompatible
with the simulations and data.

\subsection{Likelihood Validation}
\label{app:likevalid}

\begin{figure}
\resizebox{\columnwidth}{!}{\includegraphics{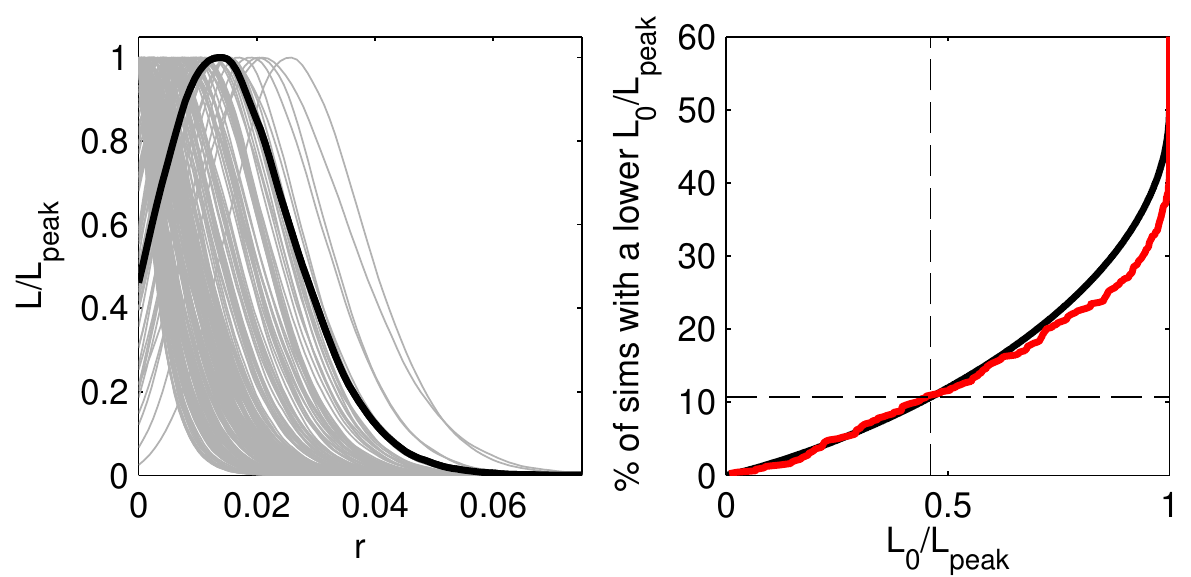}}
\caption{
{\it Left:} Likelihood curves for $r$ when running the baseline
analysis on 200 of the lensed-\lcdm+dust+noise simulations.
The real data curve is shown overplotted in heavy black.
{\it Right:} The CDF of the zero-to-peak ratio (red) of the curves shown
at right as compared to the simple analytic ansatz (solid black)
$\frac{1}{2} \left( 1-f \left( -2\log{L_0/L_{\rm peak}} \right) \right)$
where $f$ is the $\chi^2$ CDF (for one degree of freedom).
About one tenth of the simulations offer more evidence for non-zero
$r$ than the real data when the true value is actually zero (dashed black).}
\label{fig:simcons}
\end{figure}

To validate the likelihood analysis on the real data we run full
\texttt{COSMOMC} runs on the ensemble of lensed-\lcdm+dust+noise simulations.
Fig.~\ref{fig:simcons} is an update of Fig.~19 of BK15.
The left panel shows the $r$ constraint curves for the simulations,
while the right panel compares the CDF of the zero-to-peak likelihood ratios
to the simple analytic ansatz
$\frac{1}{2} \left( 1-f \left( -2\log{L_0/L_{\rm peak}} \right) \right)$
where $f$ is the $\chi^2$ CDF (for one degree of freedom).
We find that 60\% of the simulations peak at zero,
and 11\% have a lower zero-to-peak ratio than the real data---i.e.\
show more evidence for $r$ when the true value is in fact zero.

\begin{figure*}[htb]
\begin{center}
\resizebox{\textwidth}{!}{\includegraphics{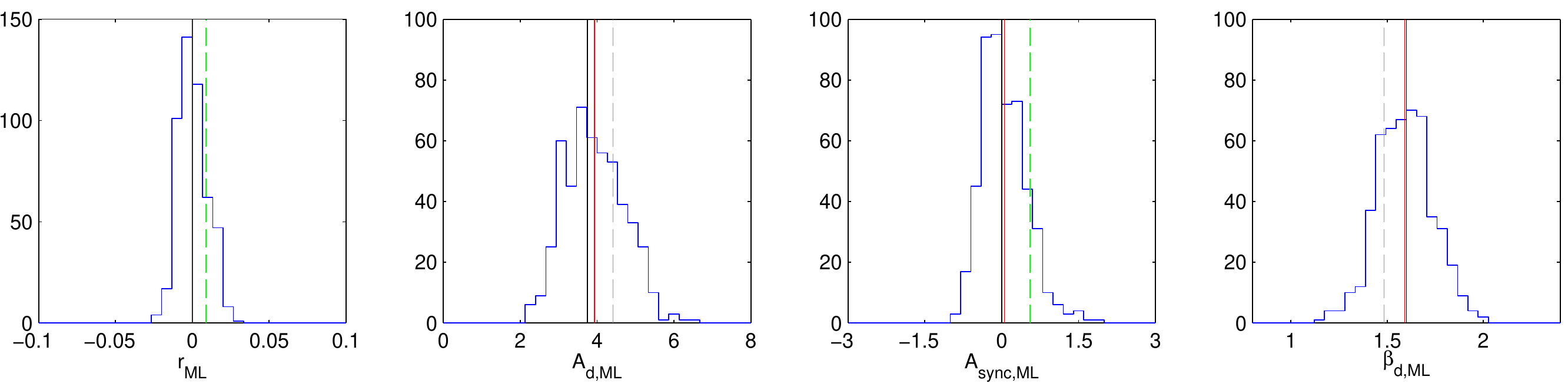}}
\end{center}
\caption{
Results of a validation test running maximum likelihood
search on the standard lensed-\lcdm+dust+noise simulations
($r=0$, $\Adf=3.75$\,\uksq, $\Bd=1.6$, $\ad=-0.4$, $\As=0$).
The blue histograms are the recovered maximum likelihood values with the
red lines marking their means and the black lines showing
the input values.
The green dashed lines are the real data values for the BK18 baseline
data set.
In the left panel $\sigma(r)=0.009$.
See Appendix~\ref{app:likevalid} for details.}
\label{fig:likevalid}
\end{figure*}

An alternate (and much faster) likelihood validation
exercise is to run maximum likelihood searches,
with non-physical parameter values allowed (such as negative $r$).
When running on simulations generated
according to the model being re-fit, we then
have an a priori expectation that the input parameter values
should be recovered in the mean.
Fig.~\ref{fig:likevalid} shows the results when running
on the standard lensed-\lcdm+dust+noise simulations.
Only the four well constrained parameters are shown,
the others being prior dominated.
The input values are recovered in the mean although formally
there is a small bias detectable on $\Ad$.
We prefer this $\sigma(r)$ measure of the intrinsic constraining
power of the experiment since it is independent of the particular
noise fluctuation that is present in the real data.

\subsection{Exploration of Alternate Foreground Models}
\label{app:altfore}

We now extend the maximum likelihood validation study to
simulations using alternate foreground models, updating
Appendix~E.4 of BK15, and giving results in Table~\ref{tab:altfgmod}.
Since the BK18 baseline analysis now has $\Bd$ free we provide
results for this case only.
Our basic set of simulations includes Gaussian
realizations of dust generated according to the parametric
model which we use when re-fitting, and these are thus expected
and to give unbiased results---as we see they do in the first row
of the table.
We also make decorrelated simulations according to our
parameterization of decorrelation, and these are thus also
expected to give unbiased results when re-fit with that extension
to the parameterization---as we see they do in the second line of
the table.
In addition we make so called ``amplitude modulated Gaussian''
dust simulations where Gaussian full sky realizations are
multiplied by the square root of maps of degree scale $BB$ power
measured from small patches of the \planck\ 353\,GHz map in a
similar manner to Fig.~8 of Ref.~\cite{planckiXXX}.
Such variation of the dust amplitude across the field can potentially
produce bias on $r$ given that our re-fit model assumes a single
$\Ad$ value applies to both the smaller \biceptwo/\keck\ field
and the larger \bicepthree\ field.
However, in practice the third line of the table shows no
detectable bias.

\newlength{\x}
\settowidth{\x}{+}
\begingroup
\squeezetable
\begin{table}[pht]
\caption{\label{tab:altfgmod}
Uncertainty and bias on $r$
in simulations using a variety of foreground models.
For the strongly decorrelated model bias is expected
when refit without a decorrelation parameter
so this case is in parentheses.
}
\begin{ruledtabular}
\begin{tabular}{l r r c c}
      & \multicolumn{1}{c}{$\overline{A_d}$} & \multicolumn{1}{c}{$\overline{A_s}$} & \multicolumn{2}{c}{$\sigma(r)$, $\overline{r}/\sigma(r)$} \\
Model & (\uksq) & (\uksq) & no decorr. & with decorr. \\
\hline
Gaussian        &  3.9 & 0.1  &   0.009, \makebox[\x] { }0.0$\sigma$  &  0.010, \makebox[\x] { }0.0$\sigma$  \\
G.\ Decorr.     &  5.1 & 0.1  &  (0.012, \makebox[\x] {+}2.1$\sigma$) &  0.014, \makebox[\x]{--}0.1$\sigma$  \\
G.\ amp.\ mod.\ &  4.4 & 0.0  &   0.009, \makebox[\x] { }0.0$\sigma$  &  0.010, \makebox[\x] { }0.0$\sigma$  \\
PySM 1          & 11.3 & 0.9  &   0.010, \makebox[\x] {+}0.1$\sigma$  &  0.012, \makebox[\x] {+}0.2$\sigma$  \\
PySM 2          & 25.6 & 0.8  &   0.011, \makebox[\x] { }0.0$\sigma$  &  0.012, \makebox[\x] { }0.0$\sigma$  \\
PySM 3          & 11.6 & 0.9  &   0.011, \makebox[\x] { }0.0$\sigma$  &  0.013, \makebox[\x]{--}0.1$\sigma$  \\
MHDv3           &  3.2 & 7.1  &   0.012, \makebox[\x]{--}0.1$\sigma$  &  0.013, \makebox[\x]{--}0.4$\sigma$  \\
MKD             &  3.9 & 0.1  &   0.009, \makebox[\x] { }0.1$\sigma$  &  0.010, \makebox[\x] { }0.0$\sigma$  \\
Vansyngel      &  5.5 & 0.1  &   0.009, \makebox[\x]{--}0.1$\sigma$  &  0.010, \makebox[\x] { }0.0$\sigma$  \\
\end{tabular}
\end{ruledtabular}
\end{table}
\endgroup

We also consider a suite of third party foreground models
which are constructed in a variety of ways and which do not necessarily
conform to any specific parametric model.
Hence they may potentially produce bias in $r$ at levels relevant compared
to the noise.
The fourth and subsequent rows of Table~\ref{tab:altfgmod} summarize
the results.
These third-party models provide only a single realization
of the foreground sky, and we add it on top of
each of the lensed-\lcdm+noise realizations that are used in
the standard simulations.
The PySM models 1, 2 and 3~\citep{thorne17} are unchanged from BK15.
They have $\Ad$ in our sky region which is much greater that the actual
level and this modestly increases $\sigma(r)$, although interestingly by an amount
that is fractionally less than the increase seen in BK15.
The MHD model~\citep{kritsuk17,kritsuk18} is also unchanged from BK15 and has a
level of synchrotron which is now in strong conflict with the data.

The last two models in Table~\ref{tab:altfgmod} are added for this paper.
The MKD model~\cite{mkd} is a three-dimensional model of polarized galactic dust emission
that takes into account the variation of the dust density, spectral index and
temperature along the line of sight, and contains randomly generated small
scale polarization fluctuations.
This model is constrained to match observed dust emission on large scales, and
match on smaller scales extrapolations of observed intensity and polarization power
spectra.
The Vansyngel model~\cite{vansyngel} is also multi-layer.
Each layer has the same intensity (constrained by the Planck intensity map), but
different magnetic field realizations.
It produces $Q/U$ by integrating along the line of sight over these multiple
layers of magnetic fields.
Neither of these models produces detectable bias on $r$.

\section{Systematics}
\label{app:syst}

Systematics effects in \bicep/\keck\ datasets have been studied in detail in
Refs.~\cite{biceptwoII,biceptwoIII,biceptwoIV,biceptwoXI}.
In this section, we assess the impact of the most important of these effects that
could bias our current constraints on $r$.
We estimate these biases by looking at the shift in maximum-likelihood parameters
derived from simulations and/or the real data.
For the case of bandpass uncertainty, we also calculate the full multi-dimensional
likelihood of the real data for a model with additional nuisance parameters, 
and compare that to the BK18 baseline result.
For all effects considered, the magnitude of the estimated bias on $r$ is
significantly smaller than our statistical uncertainty, 
and not all of the biases have the same sign, which should lead to partial
cancelation.

\subsection{Temperature-to-Polarization Leakage from Residual Beam Mismatch}
\label{app:tpleakage}

Following Ref.~\cite{s4forecast20} systematic effects may be classified
as additive, characterized by contamination in polarization maps not correctly
estimated by the noise model, or non-additive, which includes most signal
calibration errors.
The most important additive effect for our datasets is
temperature-to-polarization leakage resulting from
residual beam mismatch, crosstalk, or similar effects.
Following the methodology developed in Ref.~\cite{biceptwoXI},
we evaluate the impact of
\tp\ leakage from undeprojected main beam mismatch on $r$ recovery.
We analyze the shift in maximum likelihood $r$ estimation for a set of
499 simulations which have an added bias corresponding to an estimate of
\tp\ leakage. This leakage estimate comes from specialized
``beam map simulations'', as described in Sec.~5 of Ref.~\cite{biceptwoXI}.
We report the median and 1-$\sigma$ standard deviation of the
realization-to-realization recovered values of $r$.

\begin{figure}[t]
\resizebox{\columnwidth}{!}{\includegraphics{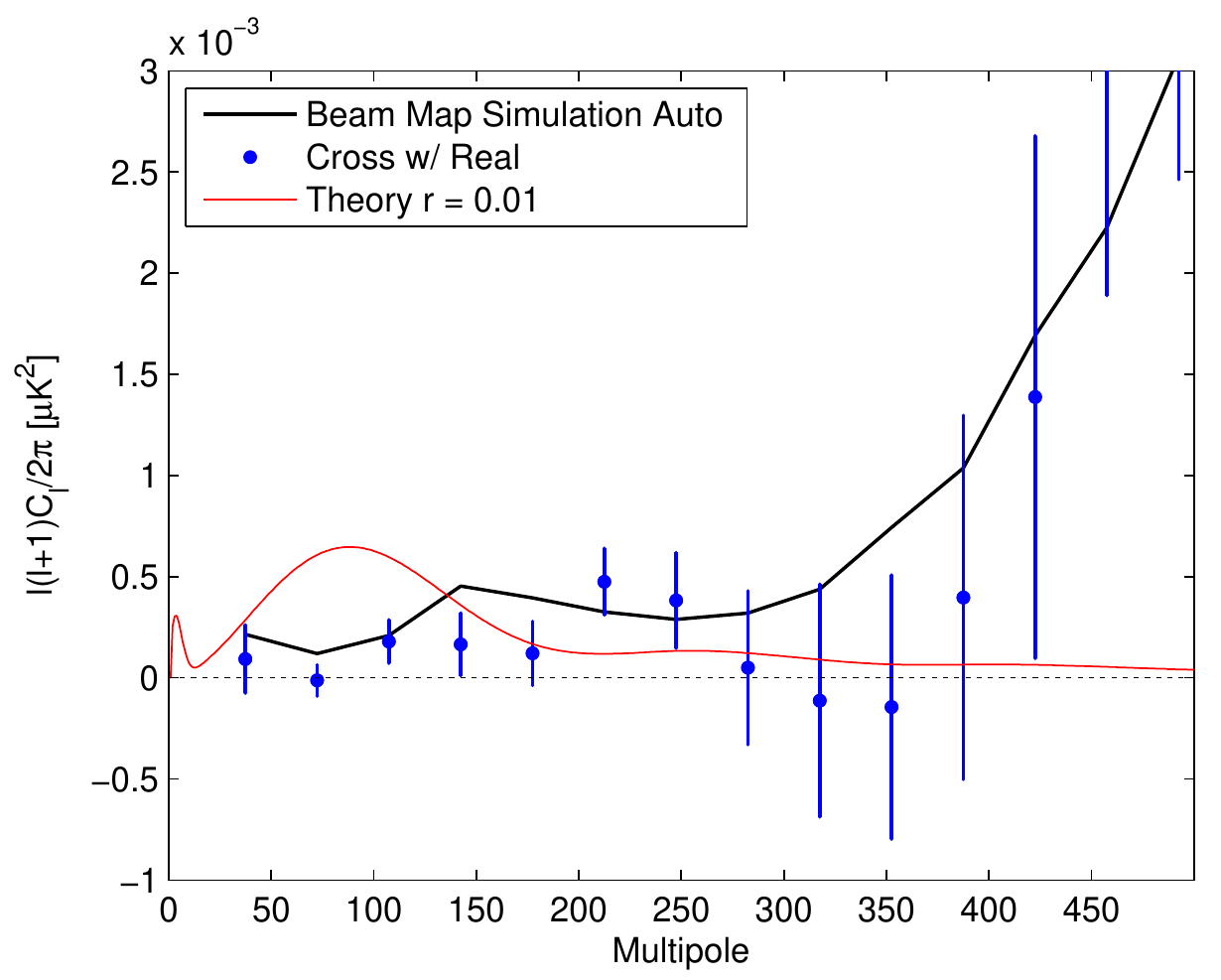}}
\caption{
$BB$ power spectra corresponding to \tp\ leakage in \bicepthree,
as predicted by beam map simulations.
The black line shows the auto-spectrum, which has been
noise debiased (from noise in the beam map measurement).
The blue circles show the cross of the beam map simulation with
real BK18 data, with error bars derived from the cross between
the fixed beam map simulation with
499 lensed-\lcdm+dust+noise simulations.}
\label{fig:bmsim_spec}
\end{figure}

Fig.~\ref{fig:bmsim_spec} shows the beam map simulation $BB$ auto spectrum and
cross with real for \bicepthree.
The corresponding spectra for the \keck\ 95\,GHz, 150\,GHz and 220\,GHz
bands were found to be very similar to the BK15 results shown in Ref.~\cite{biceptwoXI}.

We report results corresponding to the ``CMB data-driven'' scenario
presented in Ref.~\cite{biceptwoXI}, for which leakage templates are added
only on auto-frequency spectra, as the leakage templates show no evidence
of a common-mode component that would bias the cross-frequency spectra.
Specifically, we add a leakage contribution whose
mean amplitude is set by the cross spectra of beam map simulations with real
BK18 maps (blue circles in Fig.~\ref{fig:bmsim_spec}), plus a random component
with the standard deviation of the cross spectra between the beam map simulations the
standard lensed-\lcdm+dust+noise simulations (error bars
in Fig.~\ref{fig:bmsim_spec}).

The recovered bias is $\Delta(r)=1.5 \pm 1.1 \times 10^{-3}$, which is
subdominant compared to our statistical uncertainty $\sigma(r) = 0.009$.
Compared to BK15 results, the reduction in \tp\ bias
($\Delta(r) = 2.7 \pm 1.9 \times 10^{-3} \rightarrow 1.5 \pm 1.1 \times 10^{-3}$)
nearly matches the reduction in statistical uncertainty
($\sigma(r) = 0.020 \rightarrow 0.009$).
The driving factor in this improvement in the recovered bias is the
inclusion of \bicepthree, which demonstrates lower differential beam
power than \keck\ and \biceptwo, both before and after deprojection.
Note that this bias estimate naturally includes the effect
of \tp\ leakage due to crosstalk arising from our time-domain
multiplexed readout. Beam simulations with crosstalk deprojected
(as described in Ref.~\cite{biceptwoIII}) show that the contribution
to $\Delta(r)$ from crosstalk is small.

This analysis highlights the importance of taking high-fidelity far-field beam map
(FFBM) measurements \textit{in situ}, which are essential in validating the
removal of leading order difference modes via deprojection, and in
ensuring that the bias on $r$ due to undeprojected residual leakage is
properly controlled.
In this leakage analysis for BK18, care was taken to
minimize systematic contamination in the composite beam maps from the
FFBM analysis pipeline itself, e.g. by minimizing non-Gaussian noise and
excess smoothing.
Further improvements can be made in the future,
by developing improved noise estimates that accurately capture per-map pixel
uncertainty in the beam maps, especially where the coverage from the redirecting
mirror varies.
Higher S/N beam maps out to larger radii can also be
constructed by combining FFBM measurements made using the thermal chopper
with those made using a high-powered noise source~\cite{biceptwoIV}.
This allows a quantification of leakage due to
mismatch not only of the main beam, but of any region of the beam within
$<20\deg$ of the main beam.
Improvements such as these will ensure that
\tp\ leakage does not become a limiting effect as $\sigma(r)$ further
declines with the deployment of the new \biceparray\ receivers.

\subsection{Other Additive Systematics Effects}

Other additive systematic effects including EMI, magnetic, thermal,
and other scan-synchronous sources of residual contamination, and
ghost beams, were studied in Ref.~\cite{biceptwoIII} and upper limits placed
on their contributions equivalent to $r<10^{-4}$, except in the case
of EMI contamination from the satellite uplink transmitter, which was
bounded at $r<1.7 \times 10^{-3}$ with 95\% confidence.
Levels of these potential sources of contamination have been monitored
throughout collection of the current dataset (including EMI reduction
from additional shielding at the satellite uplink),
and contributions from these effects remain negligible.
More significant, but still subdominant for the current analysis, are
uncertainties from the contributions of polarized flux from point
sources in our field, which estimates suggest may yield
an upward bias on $r$ of 1--3$\times 10^{-3}$~\cite[and recent internal work]{biceptwoI,battye10},
and imperfections in our noise bias estimates as they interact with
leakage deprojection filtering, which recent preliminary work
suggests may yield a downward bias on $r$ of $2 \times 10^{-3}$.
While small compared to our present uncertainties,
we expect in future rounds of analysis to be able to mitigate both these effects.
\vspace{-12pt} 

\subsection{Bandpass Uncertainty}
\label{app:bpuncertainty}

Given the importance of multi-frequency component separation to the current result, 
the most important non-additive systematic effect is spectral bandpass uncertainty.
Bandpass parameters (center and width) for the \bicepthree\ (95\,GHz) and \keck\
(95, 150 and 220\,GHz) receivers are determined using a Fourier-Transform
Spectrometer (FTS).
We consider various sources of uncertainty in these measurements and estimate them to be small, at the 1\% level or less.
As a conservative upper limit, we estimate the impact of errors up to
2\% for the band centers of the \bicepthree\ and \keck\ data sets.

First, we simulate all possible combinations of $\pm$ 2\% bandcenter shifts on
the \bicepthree\ and \keck\ bands.
We use maximum likelihood searches on 499 lensed-\lcdm+dust+noise
simulations to derive upper bounds for biases on recovered cosmological
parameters.
We show that, for all 8 parameters that we fit for in our maximum likelihood
framework, the expected biases are low and within statistical uncertainties.
In particular, we show that the worst case scenario would result in a bias on
$r$ of $|\Delta(r)|=8.4 \pm 5 \times 10^{-4}$, well below the statistical
uncertainty $\sigma(r) = 0.009$

Secondly, we include nuisance parameters on \bicepthree\ and \keck\ bands in the
CosmoMC likelihood to marginalize over errors in the bandpass measurement.
Similar to BK15, we consider one nuisance parameter per frequency band that 
represents a fractional shift in the band center, and use a Gaussian prior 
with mean/standard deviation of (0/0.02) on each of these parameters.
Compared to the BK18 baseline, the shift in the 95\% upper limit on $r$ 
due to the addition of these nuisance parameters is $3 \times 10^{-5}$, 
negligible compared to the statistical uncertainty.
The shifts on dust parameters ($A_d$ and $\beta_d$) are a few percent, and
still within statistical uncertainty and upper bounds predicted using
maximum likelihood searches on simulations.
Additionally, we compare the shifts seen from the real data to those obtained from 21
simulations on which we run the CosmoMC likelihood analysis, with and without
bandpass nuisance parameters.
We compute the realization-to-realization shift in the likelihood peak and found
none of the shifts observed in real data for $r$ and dust parameters to be
statistically significant compared to random fluctuations in simulations.

\end{appendix}

\end{document}